\pdfoutput=1

\documentclass[referee]{raa}            

\usepackage{graphicx,times}             
\usepackage{natbib}
\usepackage{amssymb,amsmath}
\bibpunct{(}{)}{;}{a}{}{,}
\usepackage[a4paper=true,pagebackref=true]{hyperref}
\hypersetup{colorlinks = true, linkcolor = green, anchorcolor = red, citecolor = blue, filecolor = red, pagecolor = red, urlcolor = red}

\begin{document}

    \title{  The variability of the narrow-line Seyfert 1 galaxies from the Pan-STARRS's view 
}

   \volnopage{Vol.0 (20xx) No.0, 000--000}      
   \setcounter{page}{1}          

   \author{Hongtao Wang
      \inst{1}
   \and Yanping Su
      \inst{2}
  \and Xue Ge
      \inst{3}
     \and Yongyun Chen 
      \inst{4}      
   \and  Xiaoling Yu
      \inst{5}
   }

   \institute{  School of Science, Langfang Normal University, Langfang, 065000, China  ({\it e-mail: wanghongtao@lfnu.edu.cn })\\
        \and
        School of Life Science, Langfang Normal University, Langfang, 065000, China\\
        \and 
        School of Physics and Electronic Engineering, Jiangsu Second Normal University, Nanjing, Jiangsu 211200, China
        \and
        The College of Physics and Electronic Engineering, Qujing Normal University, Qujing, 655011, China \\
        \and
         School of Astronomy and Space Science, Nanjing University, Nanjing, 210093, China\\
       \vs\no
   {\small Received~~xxx month day; accepted~~xxx~~month day}}

\abstract{  By means of the data sets from the Pan-STARRAS1 survey, we have systematically examined the relationship between the variability characteristics and the physical parameters of the largest NLS1 galaxy sample up to now. The results are summarized as follows: (1). We find significant anti-correlations between variability amplitude and absolute magnitude in g,r,i,z and y bands, respectively, which are consistent with the results in previous works. (2) The correlations between the variability amplitude in optical band and many physical parameters (e.g., $\lambda L(5100\rm\AA)$, black hole mass, Eddington ratio, $R_{4570}$ and $R_{5007}$ are investigated. The results show the variability amplitude is significant anti-correlation with $L(5100\rm\AA)$, $M_{\rm BH}$, Eddington ratio and $R_{4570}$, but positive correlation with $R_{5007}$. The relation could be explained by the simple standard accretion disk model. (3) We further investigate the relationship between optical variability and radio luminosity/radio loudness. The results present weak positive correlation in g and r bands, insignificant correlation in i, z and y bands. The large error of the approximate fraction of host galaxy in i, z and y bands may lead to the insignificant correlations.      
 \keywords{AGN: NLS1 --- galaxies: evolution --- galaxies: individual }  
}

   \authorrunning{Wang et al. }            
   \titlerunning{ The variability of NLS1 galaxies }  

   \maketitle

%
%
\section{Introduction}           
\label{sect:intro}

The optical variability of active galactic nuclei (AGNs) has been found since their discovery \citep{1963ApJ...138...30M}. Irregular variability of quasars/AGNs is ubiquitous across all wavebands \citep{1997ARA&A..35..445U}. It is generally recognized that the emission in optical band originates from the thermal radiation in the accretion disk surrounding supermassive black holes (SMBHs). The typical variability amplitude usually shows less than two magnitudes within the timescale of a few months to years \citep[e.g.,  ][]{2004IAUS..222..525I,2004ApJ...601..692V}. The optical variability carries much valuable information, which could be used to trace the physical processes around the SMBHs \citep{2017ApJ...847..132G}. 
 
The optical variation on the order of hours to years is one of the main features of active galactic nuclei. The optical variability of quasars/AGNs is irregular and complicated. It is commonly believed that the emission in optical band originates from the optically thick accretion disk driven by the central supermassive black hole, but the physical processes producing its variability are not clearly  understood. However, the statistical study of the variability sample on fundamental physical parameters provides a new view for us to uncover the variability mechanisms. Many investigations of AGN/quasars indicated that the variability was correlated with wavelength 
\citep[e.g.,][]{1996ApJ...463..466D,2001AJ....121.1872H}, luminosity \citep[e.g.,][]{2000A&AS..143..465H,2019ApJ...877...23L}, time lag \citep[e.g.,][]{2005AJ....129..615D}, and redshift  \citep[e.g.,][]{2004ApJ...601..692V,2011A&A...525A..37M,2011A&A...527A..15W}. Some results showed the variability was correlated with $\rm M_{BH}$  \citep[e.g.,][]{2009ApJ...696.1241B,2012ApJ...758..104Z,2019ApJ...877...23L}, and anti-correlated with the Eddington ratio  \citep[e.g.,][]{ 2010ApJ...716L..31A,2010ApJ...721.1014M,2019ApJ...877...23L}.  

Seyfert galaxies are a subclass of AGNs, which are generally classified into Seyfert 1 and Seyfert 2 galaxies. Seyfert 1 galaxies usually present the broad emission line of a few thousand $km~s^{-1}$ originating from the broad line region (BLR). Seyfert II galaxies show the emission line less than 1000 $km~s^{-1} $ from the narrow line region (NLR). Narrow-line Seyfert 1 (NLS1) galaxies have narrow $H\beta$ line with the full width at half maximum (FWHM) $< 2000~km~s^{-1}$ and weak [O III] emission line with the flux ratio F([O III])/F(H$\beta) < 3$ (Osterbrock \& Pogge 1985). Some results show that NLS1 galaxies have smaller black hole mass $\sim10^7 M_{\odot}$ and higher accretion rate \citep{2012AJ....143...83X}. However, other results show the smaller black hole mass may be arising from the effect of the inclination angle of BLR \citep{2008MNRAS.386L..15D}.  

 A few variability studies on NLS1 galaxies are investigated previously. Only a few objects have much longer light curves, e.g., the optical variability  \citep[e.g.,][]{2006ARep...50..708D,2012ApJS..202...10S} and the X-ray variability of Ark 564 \citep{2002A&A...391..875G}, the variability of 113 bright soft X-ray AGNs  \citep{2001A&A...367..470G}. The above results show the NLS1 galaxies usually have low  variability. Based on the optical variability of six NLS1 galaxies,  \citet{2004ApJ...609...69K} found that NLS1 galaxies were less variable than broad line Seyfert 1 (BLS1) galaxies on long timescales in optical band which could be explained by the negative correlation between variability amplitude and Eddington ratio, if NLS1  galaxies actually had high accretion rate. \citet{2010ApJ...716L..31A} analyzed the  variability amplitude  ($\sigma_d$) of 58 NLS1 and 217 BLS1 galaxies by the multi-epoch photometric data sets in Stripe 82 from SDSS. The results showed significant and robust negative correlation between $\sigma_d$ and $\lambda_{Edd}$.  \citet{2013AJ....145...90A} further investigated the variability by ensemble method. In their results, the majority of NLS1 galaxies presented significant variability on the timescale larger than 10 days, but smaller variability amplitude  compared to BLS1 galaxies. In the timescale of less than 10 days, NLS1 galaxies may have different variability mechanisms comparing with BLS1 galaxies. Such as X-ray reprocessing may appear in BLS1 AGNs, however, not occur in NLS1 AGNs. Compared with the broad line counterparts, the long-term optical  variability of NLS1 galaxies had been investigated by \citet{2017ApJ...842...96R}, in which a large number of objects were analyzed by the catalogue of  \citet{2017ApJS..229...39R} with the optical data from the Catalina Real Time Transient Survey. They found that NLS1 galaxies as a class show lower variability amplitude than their broad-line counterparts. In addition to the long-term optical variability, the characteristics of intra-night optical variability in different categories of NLS1 galaxies are also well studied  \citep{2013ApJ...762..124M,2013MNRAS.428.2450P,2017MNRAS.466.2679K}. In summary, the statistical regularity of optical variability in NLS1 galaxies is relatively scarce and should be further investigated.  
 
 The Panoramic Survey Telescope and Rapid Response System (Pan-STARRS, the more details in Section 2) was operated in 2014. The photometric uncertainties of Pan-STARRS is lower than that in  PTF/iPTF, ZTF and CRTS. The other basic parameters are listed in Table 1, which show much deeper observation than that in PTF/iPTF, ZTF and CRTS. The cadence is higher than that in SDSS, PTF/iPTF, ZTF and CRTS. 
 Comparing with CRTS in \citet{2017ApJ...842...96R},  there are five photometric bands with much smaller error from Pan-STARRS, and the 5 sigma limiting magnitude $\sim$21-23 mag of Pan-STARRS which is much deeper than 19 mag in CRTS. The cadence of $\sim12$ epochs of Pan-STARRS is much higher than $1\sim4$ epochs of CRTS.  
 The sample provided by \citet{2017ApJS..229...39R} is the largest one, which contains 11 101 NLS1 galaxies and was suitable for us to investigate the variability characteristics. The Pan-STARRS with the data sets of small photometric error, high cadence and long time baseline, gives us a new chance to further investigate the variability  characteristics of NLS1 galaxies. 
 
 The paper is organized as follows. We describe the data sets and sample selection in Section 2. In Section 3, we introduce the analytical method. The results and discussions are presented in Section 4. The conclusions are given in Section 5.   
 
 \begin{table*}
\centering
\caption{ The basic parameter information of Pan-STARRS, CRTS, PTF/iPTF, ZTF and SDSS Stripe 82. } \label{Tab.XXX1} 
\resizebox{\textwidth}{15mm}{ 
\begin{tabular}{|c|c|c|c|c|c|c|}  
\hline \hline 
  & Observation band & 5$\sigma$ depth /mag  & Time baseline  &  Cadence    \\ 
\hline  
  SDSS Stripe 82  & u, g, r, i, z   & 23.9, 25.1, 24.6, 24.1, 22.8   & 1998- 2007 & $\sim$1 epoch/month \\
  CRTS/CSS,MLS    &   V     &  $\sim$  20             &  2003–2016    & 1$\sim$4 times/month  \\
  CRTS/SSS        &   V     &  $\sim$  19             &  2005 - 2013  & 1$\sim$4 times/month      \\
  PTF             &  R,g    &   R(20.5), g(21.0)         &  2009-2012    &  a 5 day cadence   \\
  iPTF            &  R,g    &  R(20.5), g(21.0)          &  2013 -2017   &  a 5 day cadence        \\
  ZTF             &  g,r,i  &  g(21.0), r(20.4),i(20.5)   &  Nov,2017- now  &  a 3 day cadence   \\
  Pan-STARRS      &  g,r,i,z,y &  23.3, 23.2, 23.1, 22.3, 21.3 & 2010-2014  & $\sim$12 epochs/month  \\
\hline  
\end{tabular}}
\end{table*} 
 
 \section{Data} 
 In this work, we have taken the 11101 NLS1 galaxies sample from the catalogue of  \citet{2017ApJS..229...39R}, which were based on the spectroscopic database of the Sloan Digital Sky Survey Data Release 12 (SDSS DR12;  \citet{2015ApJS..219...12A}).  

The optical data sets are from the Pan-STARRS operated by the institute of astronomy at Hawaii university. The facility has a wide-field astronomical imaging system. Pan-STARRS1 (PS1) is the result in the first stage of Pan-STARRS. It was completed and the database was released in Data Releases 1 and 2 (DR1 and DR2). 1.8-meter telescope with a 1.4 Gigapixel camera was used to image the sky in five broadband filters (g, r, i, z and y) during the PS1 survey, which covers the region of the north of declination -30 degree. The PS1 Science Mission started operation in March 2014, and the PS1 DR2 occurred on January 28, 2019.  
We cross match the NLS1 galaxies sample with the Pan-STARRS1 database by the matching radius of 3 arcseconds, then obtain the light curve of 11 101 samples. The maximum, minimum and the median value of the time baseline is 5.6 years, 1 day and 3.0$\sim$ 3.7 years in these five bands, respectively. The maximum, minimum and median value of the epochs number is 134, 2, and 12$\sim$20, respectively. Because the AGN variability amplitude is strongly related to the observed time interval, we select the time baseline longer than two years, thus 8078, 8955, 9318, 8777 and 7343 objects are left in g,r,i,z and y bands, respectively. The mechanism of the radio loud NLS1 galaxies in optical band originates from the relativistic jet, thus the impact of the radio loud objects should be eliminated. We cross match the sample with the Faint Images of the Radio Sky at Twenty centimeters (FIRST) Survey, 555 objects are found. Considering the variability amplitude should be larger than zero, 47, 39, 29, 24 and 18 objects are found to be radio loud objects (the radio loudness above 10) in g, r, i, z and y bands, respectively. We eliminate these objects. Ultimately, 8031, 8916, 9289, 8753 and 7325 objects are left to be radio quiet NLS1 galaxies in g,r,i,z and y bands, respectively.  
 The data information of the time baseline and epochs number in g,r,i,z and y bands is shown in Table 2. 
\begin{table*} 
\centering
\caption{ The data information in the NLS1 galaxy sample. } \label{Tab.XXX1}
\resizebox{\textwidth}{15mm}{
\begin{tabular}{|c|c|c|c|c|c|c|} 
\hline \hline
  & epoch number & epoch number & epoch number &  time baseline/day & time baseline/day  & time baseline/day \\ 
  Band &  maximum  &  mimimum  &  median value  &  maximum  & minimum  &  median value    \\
\hline
  g  &  70    &  5    &  12   &  1607.7  & 730.1  &  1096.0      \\
  r  &  78    &  5    &  14   &  1899.8  & 730.0  &  1176.8     \\
  i  &  134   &  5    &  20   &  1925.7  & 730.1  &  1503.0      \\
  z  &  54    &  5    &  13   &  2070.4  & 730.0  &  1363.1      \\
  y  &  70    &  5    &  12   &  1970.6  & 730.9  &  1428.0      \\
\hline
\end{tabular}}
\end{table*} 

\section{Methods}\label{sec:methods}
By means of the variance of observed magnitudes, we obtain the variability amplitude which is revised the contribution from the measurement errors in the data sets \citep{2007AJ....134.2236S,2017ApJ...842...96R,2019MNRAS.483.2362R}. 
The expression is as follows,  
\begin{equation} 
\Sigma=\sqrt{\frac{1}{N-1}\sum_{i=1}^{N}(m_i - <m>)^2},
\end{equation}
\noindent 
where $m_i$ is the i-th magnitude and $<m>$ is the average magnitude in the data sets. The error   $\epsilon$ is expressed as 
\begin{equation}
\epsilon^2=\frac{1}{N}\sum_{i=i}^{N}{\epsilon_{i}^{2} }.
\end{equation}
\noindent
in which $\epsilon_i$ is the i-th error. At last, the expression of the variability amplitude  is 
\begin{equation}
\sigma_m  =
  \begin{cases}
    \sqrt{\Sigma^2 - \epsilon^2},  & \quad \text{if } \Sigma>\epsilon,\\
     0,                            & \quad  \text{otherwise.}\\
  \end{cases}
\end{equation}

 Compared with other methods, it calculates the variability amplitude directly with the  model-independent method, and also takes the impact of photometric errors into account. 
 
 Considering the starlight contamination from the host galaxy of NLS1 galaxies is not negligible, we revise it by the method of \citet{2011ApJS..194...45S} in which provided us an empirical fitting formula of the average host contamination, $\dfrac{L_{5100,host}}{L_{5100,QSO}} = 0.8052- 1.5502x+0.9121x^2-0.1577x^3 $ for  $x+44 \equiv log(L_{5100,total} / erg s^{-1}) < 45.053$. We extend it to $x < 0$ due to the low luminosity of NLS1 galaxies. No correction is needed for luminosities above this value. In the formula, the rest-frame luminosity at 5100 \AA is from the catalog of NLS1 sample. The distribution of luminosity in 5100 \AA, the fractional host contamination is shown in the left and middle panel of Figure 1. The histogram distribution of variability amplitude in r band is shown in the right panel of Figure 1. The green histogram is revised by the host galaxy contamination and the blue one is not revised. The median value of the variability amplitude is  0.149 mag, 0.124 mag, 0.113 mag, 0.123 mag and 0.178 mag in g,r,i,z and y band, respectively. After correcting the host contamination, the median variability amplitude is 0.397 mag, 0.392 mag, 0.391 mag, 0.390 mag and 0.408 mag, respectively.

 \begin{figure*} 
\centering 
\includegraphics[width=0.3\linewidth]{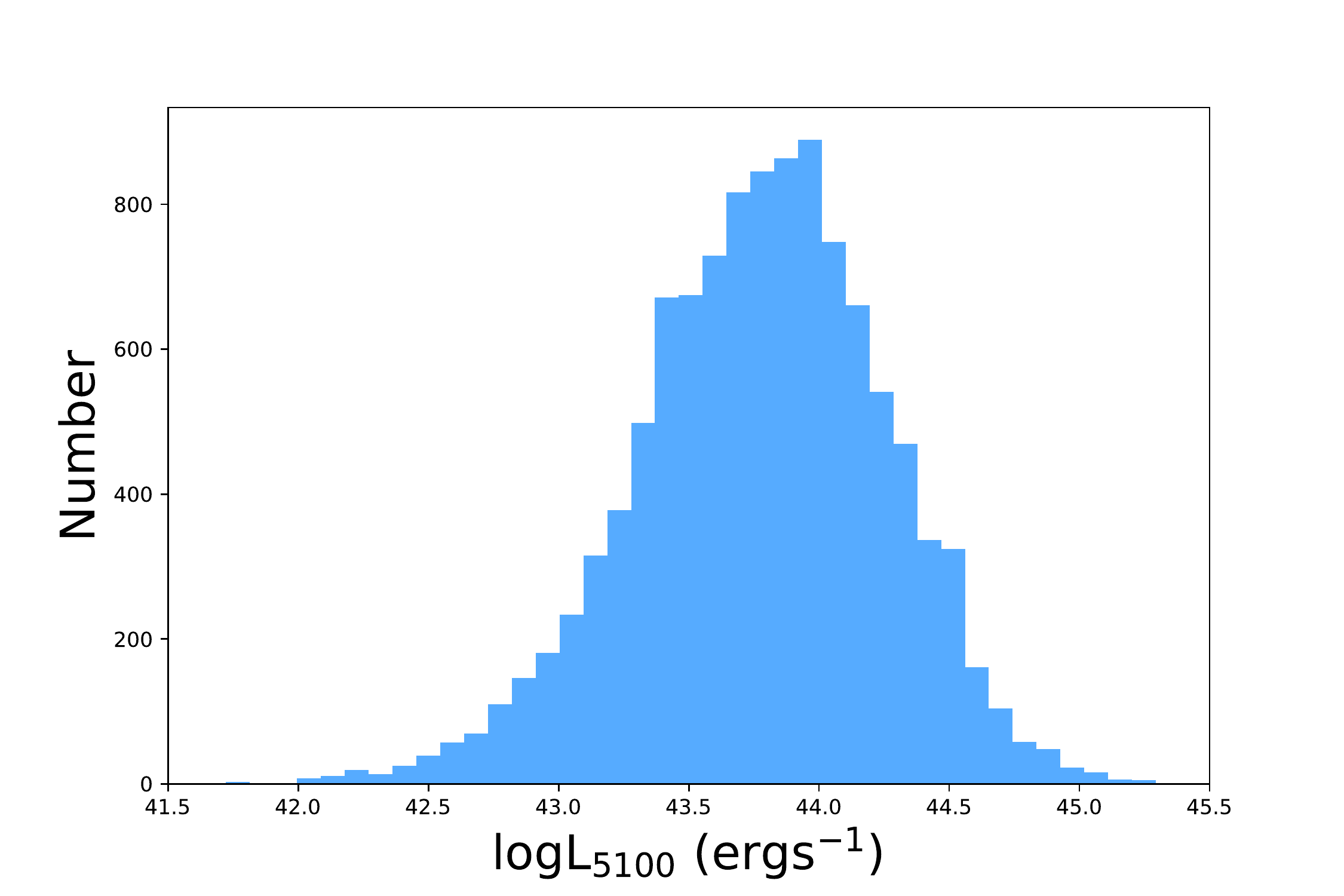}
\includegraphics[width=0.3\linewidth]{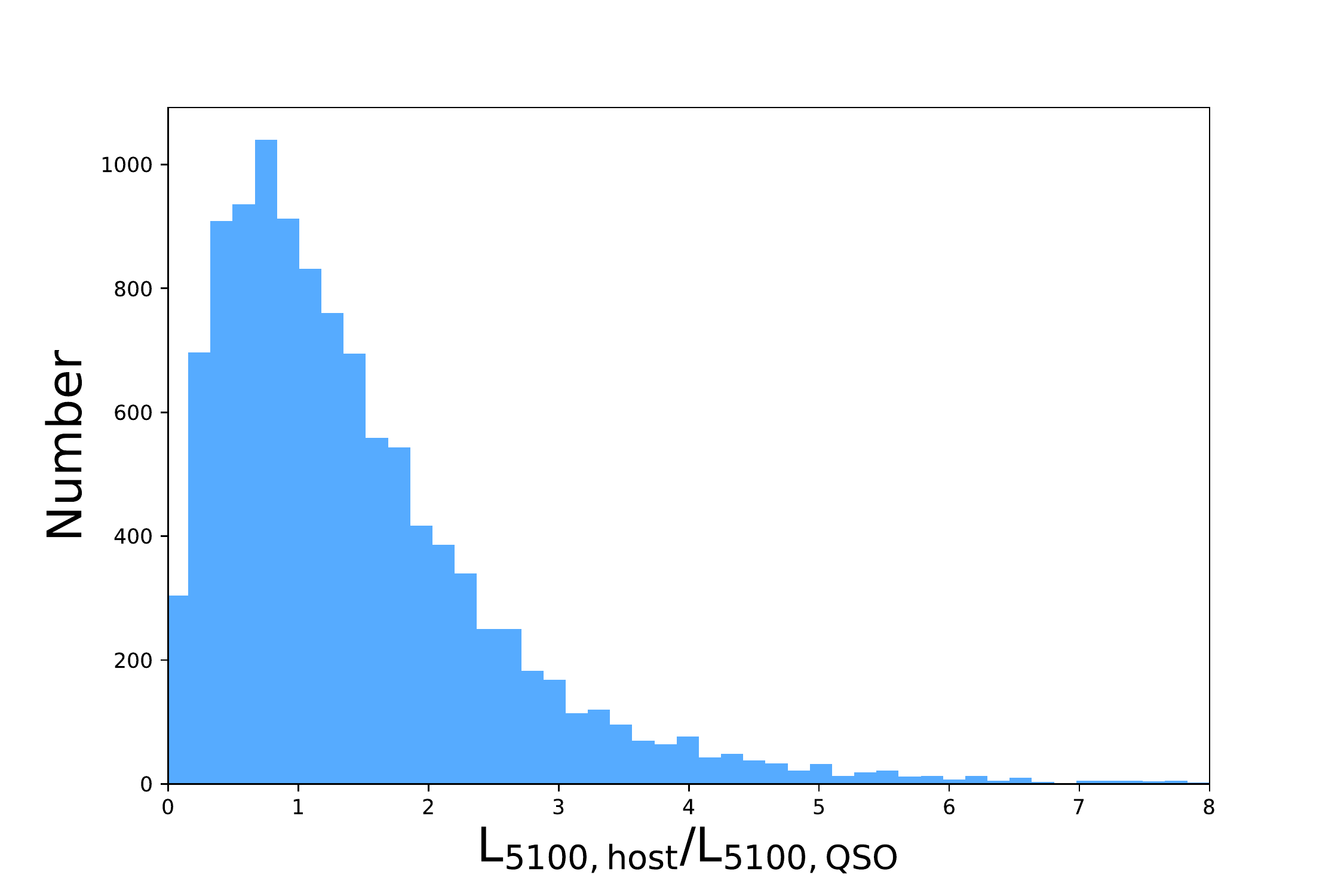}
\includegraphics[width=0.3\linewidth]{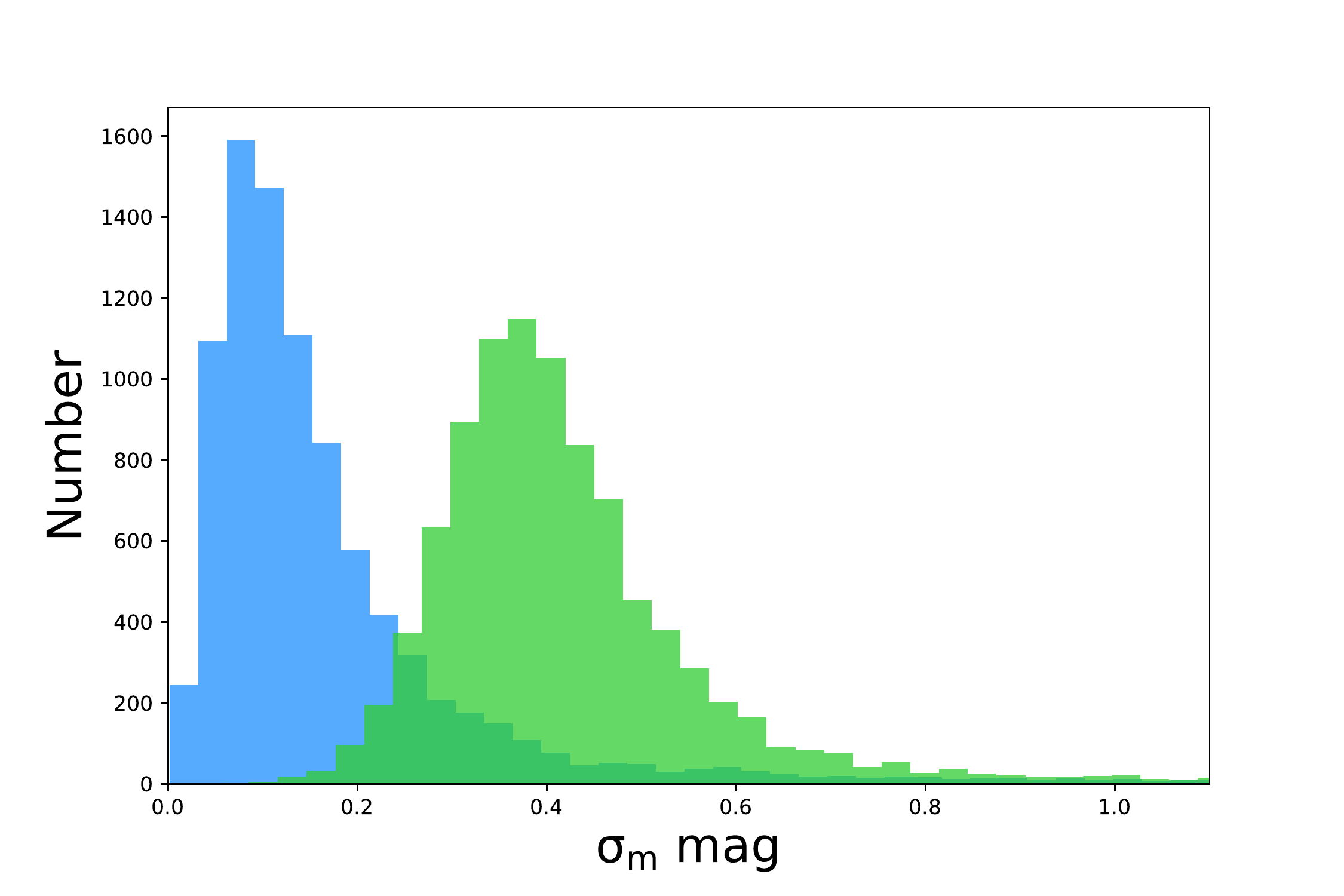}
\caption{ The distrbution of luminosity at 5100 \AA, the fractional host contamination at 5100 \AA and the variability amplitude in r band (The green histogram is revised by the host galaxy contamination and the blue one is not revised ) for NLS1 galaxies sample. }
\label{fig:linescan}
\end{figure*}
 

\section{ Results and Discussions  }
 \subsection{ The relationship between variability amplitude and absolute magnitude in g,r,i,z and y bands } 
In the section, we investigate the relationship between variability amplitude and absolute magnitude in g,r,i,z and y bands, respectively. The variability amplitude is calculated by the expression (3). The absolute magnitude reflects the intrinsic luminosity of the source. Because of the intrinsic variation, we adopt the mean value of the absolute magnitude during the time span. The results are shown in Figure 2 and Table 3. Our results show weak anti-correlations.  
An anti-correlated relationship between variability and luminosity was found in previous works adopting various AGN samples \citep{2004ApJ...601..692V,2008MNRAS.383.1232W,2012ApJ...758..104Z}. Most of them are focused on the quasar sample, and the variability work about NLS1 galaxy sample is sparse. The weak correlation in our results is consistent with the anti-correlation in quasar sample, which indicates the mechanism of NLS1 galaxy in optical band may be similar with quasar.   
\begin{figure*} 
\centering 
\includegraphics[width=0.25\linewidth]{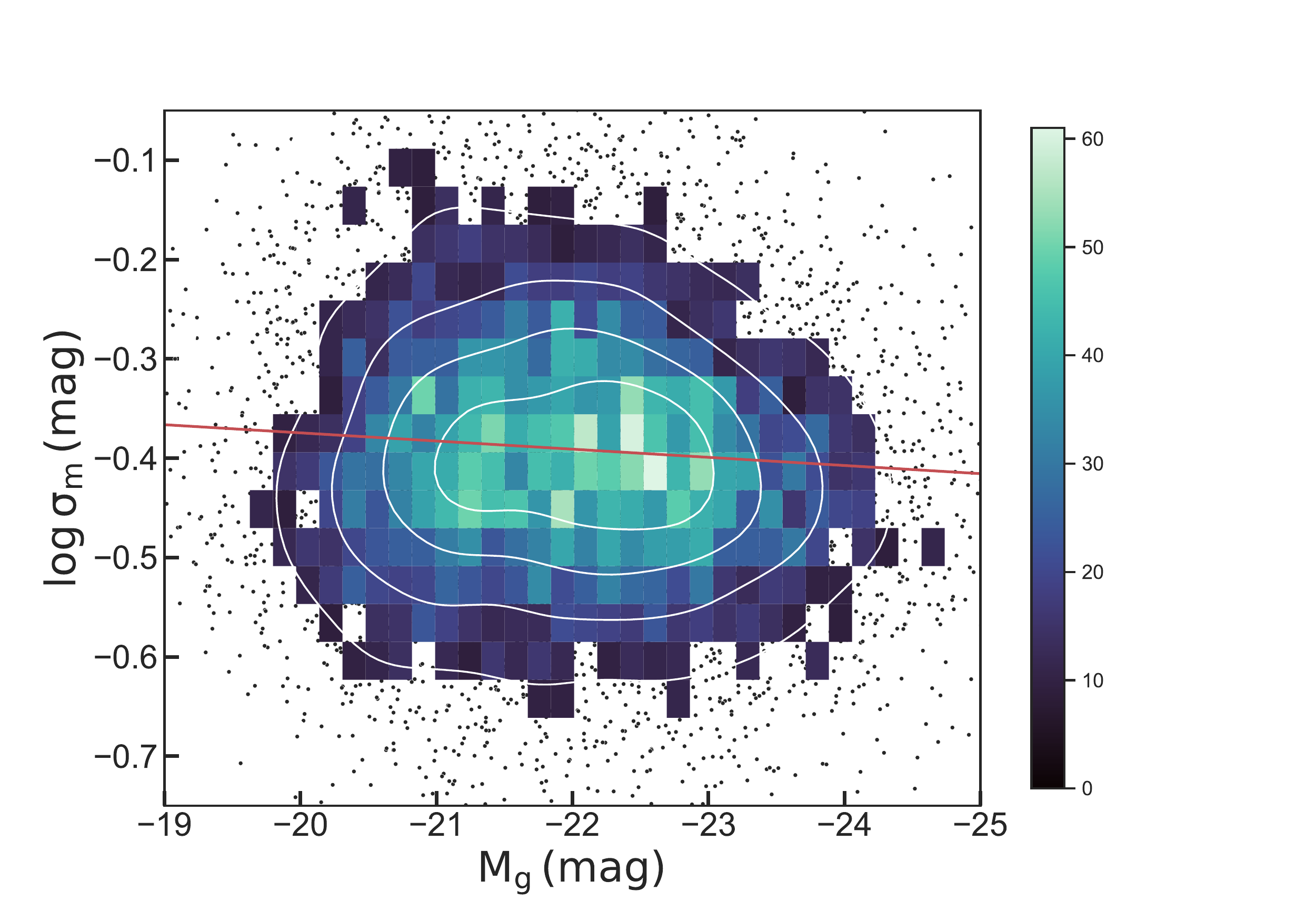}
\includegraphics[width=0.25\linewidth]{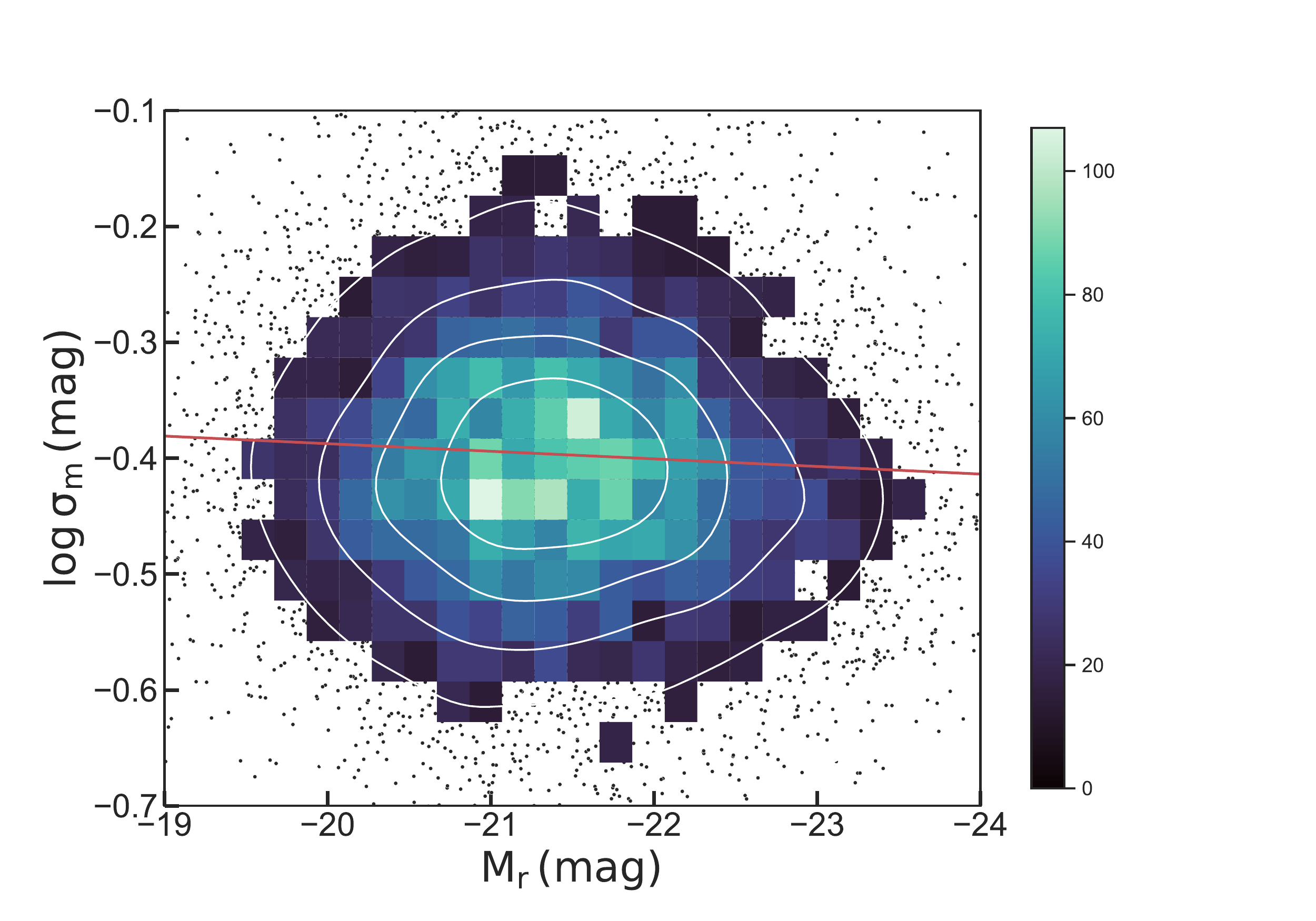}
\includegraphics[width=0.25\linewidth]{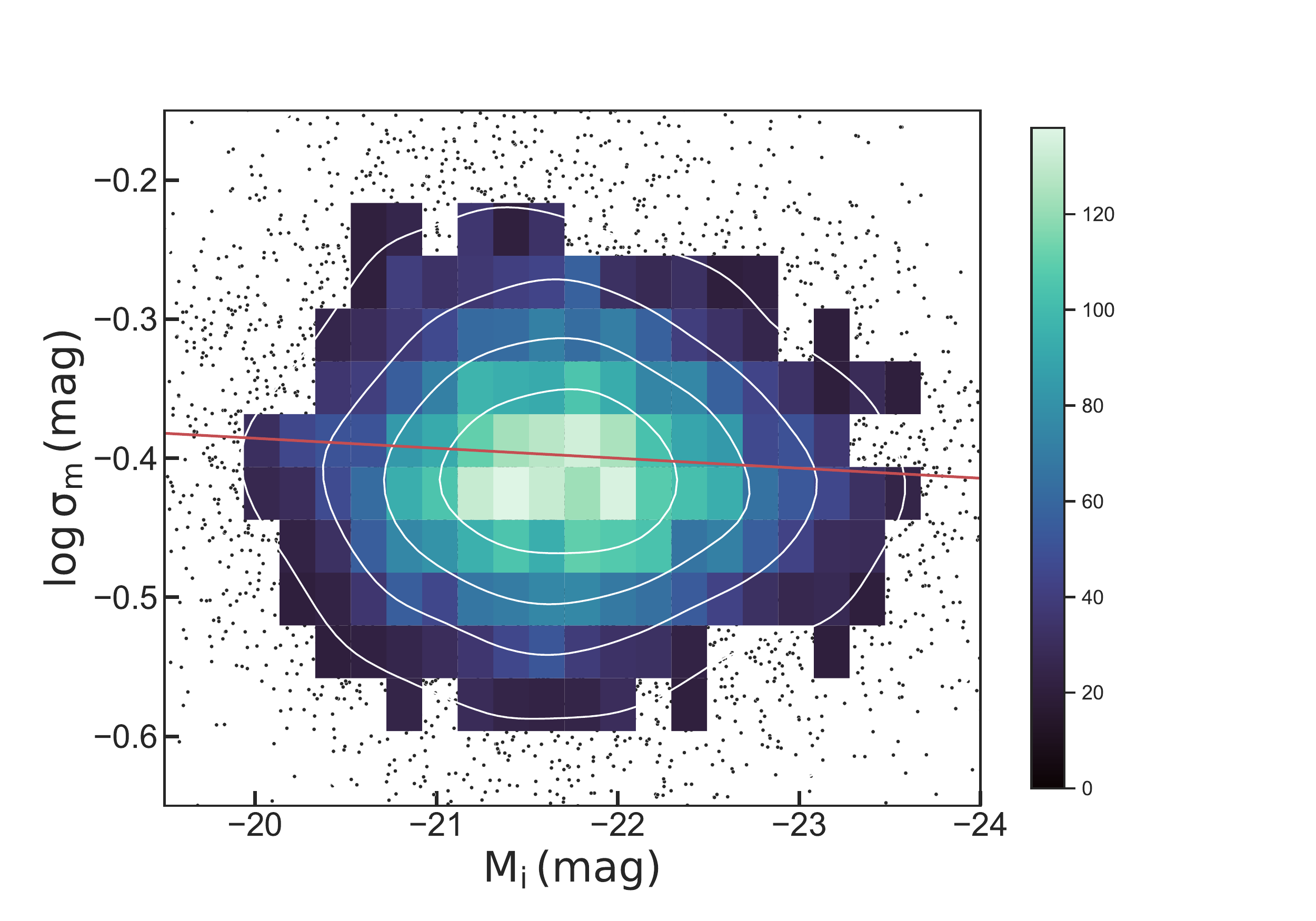}
\includegraphics[width=0.25\linewidth]{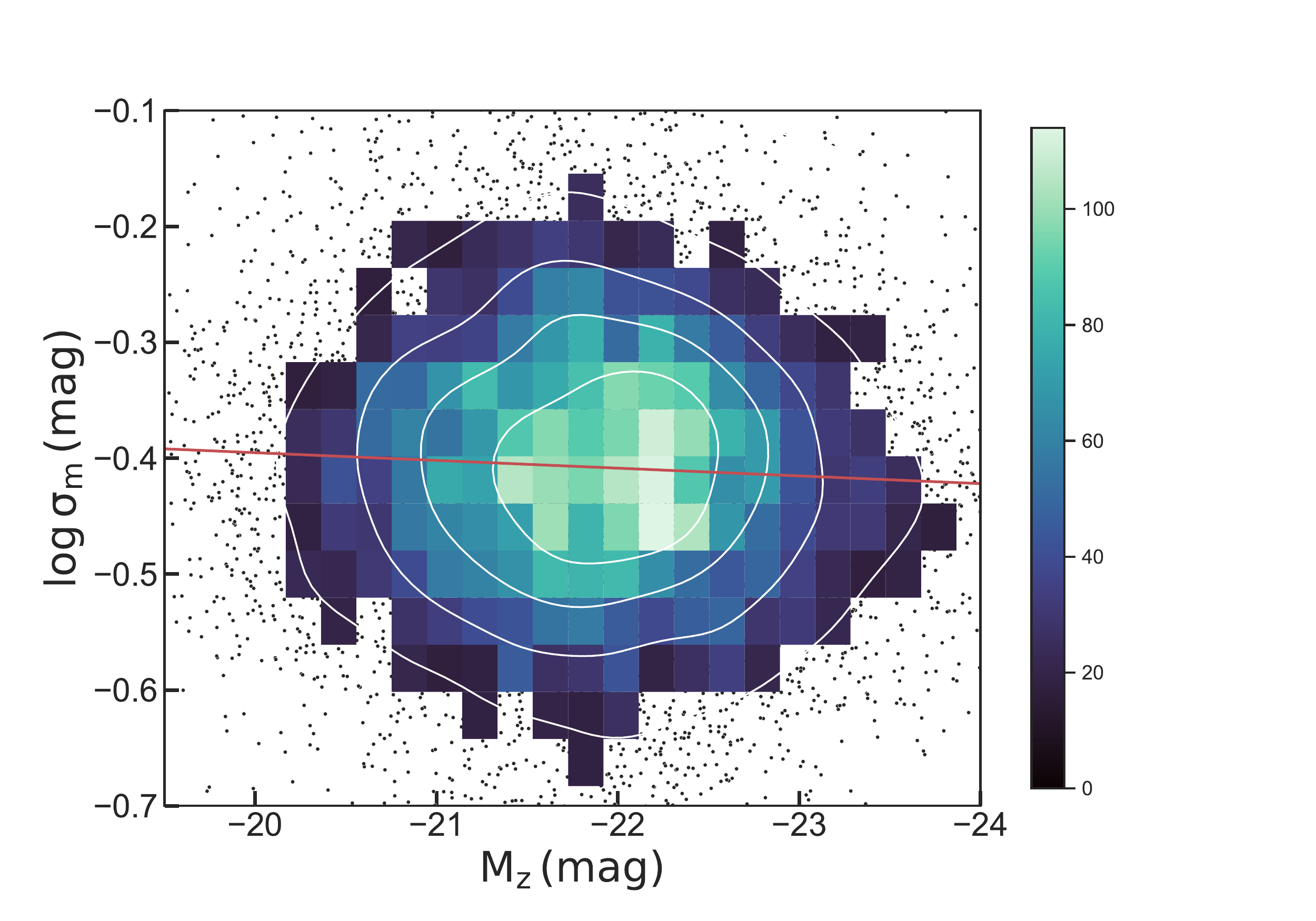}
\includegraphics[width=0.25\linewidth]{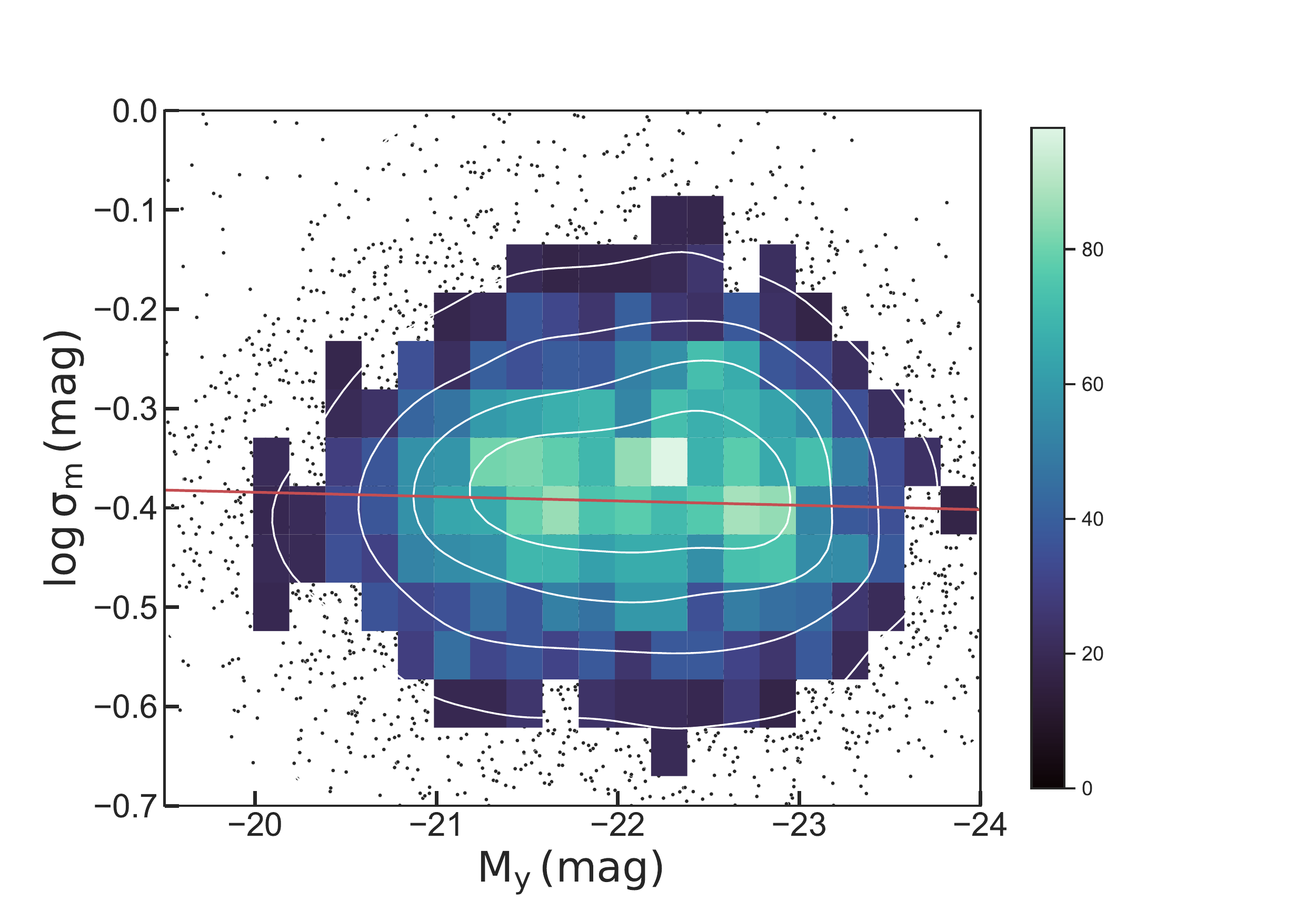}
\caption{ The relation between variability amplitude $\sigma_m$ and absolute magnitude in g,r,i,z and y bands for NLS1 galaxies sample. }
\label{fig:linescan}
\end{figure*}

\begin{table*}
\caption{ The relation between variability amplitude $\sigma_m$ and luminosity in the NLS1 galaxy sample.  } \label{Tab.XXX1}
\centering
\begin{tabular}{|c|c|c|c|}
\hline \hline
  & { Spearman correlation coefficient } & { p value } & {  expression  }   \\
\hline
 $\rm\sigma_m-g  $  &  0.064   & $6.184\times10^{-20}$     &  $-0.010x - 0.210 $   \\
 $\rm\sigma_m-r  $  &  0.042   & $4.969\times 10^{-5}$     &  $-0.065x  - 0.256 $   \\
 $\rm\sigma_m-i  $  &  0.055   & $4.294\times 10^{-10}$    &  $-0.007x - 0.242 $   \\
 $\rm\sigma_m-z  $  &  0.040   & $2.051\times 10^{-4}$    &   $-0.007x - 0.262 $   \\
 $\rm\sigma_m-y  $  &  0.022  & $2.069\times 10^{-4}$    &   $-0.004x - 0.297 $   \\
\hline
\end{tabular}
\end{table*}

\subsection{ The relation between variability amplitude and luminosity in 5100 \AA, black hole mass, Eddington ratio, $R_{4570}$ as well as $R_{5007}$ }  

In the section, we investigate the possible relationship between variability amplitude (g,r,i,z and y bands) and luminosity in 5100 $\rm\AA$ ($\lambda L_{5100}$), black hole mass $M_{BH}$, Eddington ratio $\xi_{Edd} $, $R_{4570}$ and $R_{5007}$. By matching with the variability amplitude, 6299 objects are found with $\sigma_m > 0$. The analyzing results in i band are listed in Figure 3. The Spearman coefficient, p value of no correlation and fitting expression are listed in Table 4.  

\subsubsection{ The relation between variability amplitude and luminosity in 5100 $\rm\AA$ }
We further investigate the relationship between variability amplitude and luminosity in 5100 $\rm\AA$. The variability amplitude $\sigma_m$ is calculated by the method in Section 3. The luminosity is from \citet{2017ApJS..229...39R}. The relation between variability amplitude and luminosity in 5100 $\rm\AA$ is shown in the upper left panel of Figure 3. The red line is the fitting result by the least square method. A weak anti-correlation is shown between variability amplitude and luminosity in 5100 $\rm\AA$, which is similar to the results in section 4.1.    

\citet{2010ApJ...716L..31A} analyzed the multi-epoch photometric data sets of 58 NLS1 and 217 BLS1 AGNs from the Sloan Digital Sky Survey (SDSS) in Stripe 82 region, and found the correlation between variability and luminosity is not significant. \citet{2017ApJ...842...96R} found the results presented anti-correlation in 11 101 NLS1 galaxies between variability amplitude and luminosity in 5100 $\rm\AA$ for Catalina Real Time Transient Survey(CRTS) with 5$\sim$9 years data sets and the minimum of 50 epochs. Comparing with \citet{2010ApJ...716L..31A} and \citet{2017ApJ...842...96R},  we further verify the relation of anti-correlation from the Pan-STARRS. 

\subsubsection{ The correlation analysis between variability amplitude and black hole mass } 
 We investigate the relationship between variability amplitude and black hole mass. The black hole mass is from \citet{2017ApJS..229...39R}, in which they calculated them by the virtual motion of BLR clouds. A weak anti-correlation is shown in the upper middle panel of Figure 3. Based on the data sets of 11 101 NLS1 galaxies from CRTS spanning 5$\sim$9 years, \citet{2017ApJ...842...96R}) found positive correlation in the $\sigma_d-M_{BH}$ relation, which was also found in \citet{2010ApJ...716L..31A}. But \citet{2010ApJ...716L..31A} further found the correlation disappeared when the dependency of $\lambda_{Edd}$ was considered.  \citet{2009ApJ...698..895K} found the amplitude of the short-timescale variations is significantly anti-correlated with black hole mass and luminosity with a sample of optical light curves for 100 quasars. They interpreted the optical flux fluctuations as resulting from thermal fluctuations that were driven by an underlying stochastic process, such as a turbulent magnetic field.  
 
   
 \subsubsection{ The relationship between variability amplitude and Eddington ratio } 
 The Eddington ratio $\xi_{Edd}$ is commonly considered to be the main driver of optical variability which is anti-correlated with the variability amplitude $\sigma_m$ in optical band. The Eddington ratio is estimated by   
 $\xi_{Edd} = L_{bol}/L_{Edd} $, in which $L_{bol} = 9\times \lambda L_{\lambda}(5100 \rm\AA)~\rm erg~s^{-1} $ and  $L_{Edd}=1.3\times 10^{38}M_{BH}/M_{\odot}~\rm erg~s^{-1}$(\citet{2000ApJ...533..631K}). 
  The correlation between variability amplitude and Eddington ratio is presented in the upper right panel of Figure 3. The results show a weak anti-correlation which is similar with  \citet{2010ApJ...716L..31A} and \citet{2017ApJ...842...96R}.    
 \citet{2017ApJ...842...96R} analyzed the variability of 11101 NLS1 galaxies sample and found  anti-correlated relation between variability amplitude $\sigma_m$ and Eddington ratio $\xi_{Edd}$, which may be due to the uncertainties in the calculation of $M_{BH}$ and $L_{Edd}$. \citet{2010ApJ...716L..31A} found only marginal anti-corrrelation of the NLS1 sample by the multi-epoch data sets of SDSS. 
 The anti-correlated relationship between optical variability and Eddington ratio has also been reported by many authors on the timescale of several months \citep{2013ApJ...779..187K} and several years 
\citep{2008MNRAS.383.1232W,2009ApJ...696.1241B,2010ApJ...721.1014M,2012ApJ...758..104Z,2013A&A...560A.104M} which can be understood from the simple standard accretion disk model \citep{1973A&A....24..337S}. 
If the emission originates from the inner accretion disk, the emission decreases as it propagates outward. As the Eddington ratio increases, the radius of the emission region at a given wavelength moves outward. The radius increases with the Eddington ratio since 
$ r\sim T^{-1}\sim({ \dot{m}/M_{BH})^{1/3}\lambda^{4/3}} $, where $T$ is the temperature of the disk, $\lambda$ is the wavelength, and $\dot{m}$ is the mass accretion rate. Hence the variability amplitude $\sigma_m$ decreases with the Eddington ratio increases. 

\subsubsection{ The correlation analysis between variability amplitude and $R_{4570}$ as well as $R_{5007}$ } 
In the section, we investigate the relationship between variability amplitude and $R_{4570}$ as well as $R_{5007}$. The Fe II strength $ R_{4570}$ is defined by the flux ratio of Fe II($\lambda4434–4684$) to H$\beta_b$ line, and $R_{5007}$ is calculated by the flux ratio of O[III] line to H$\beta_{tot}$ line, which are taken from  \citet{2017ApJS..229...39R}. The O[III] line originates from the narrow line region, while Fe II and H$\beta$ lines come from the broad line region. The variability amplitude is anti-correlated with $R_{4570}$ in the lower left panel of Figure 3, but positively correlated with $R_{4570}$ in the lower right panel of Figure 3, which is consistent with the results in \citet{2017ApJ...842...96R} and  \citet{2017ApJ...842...96R}) as $R_{4570}$ is related to the Eddington ratio.          

\begin{figure*}
\centering
\includegraphics[width=0.25\linewidth]{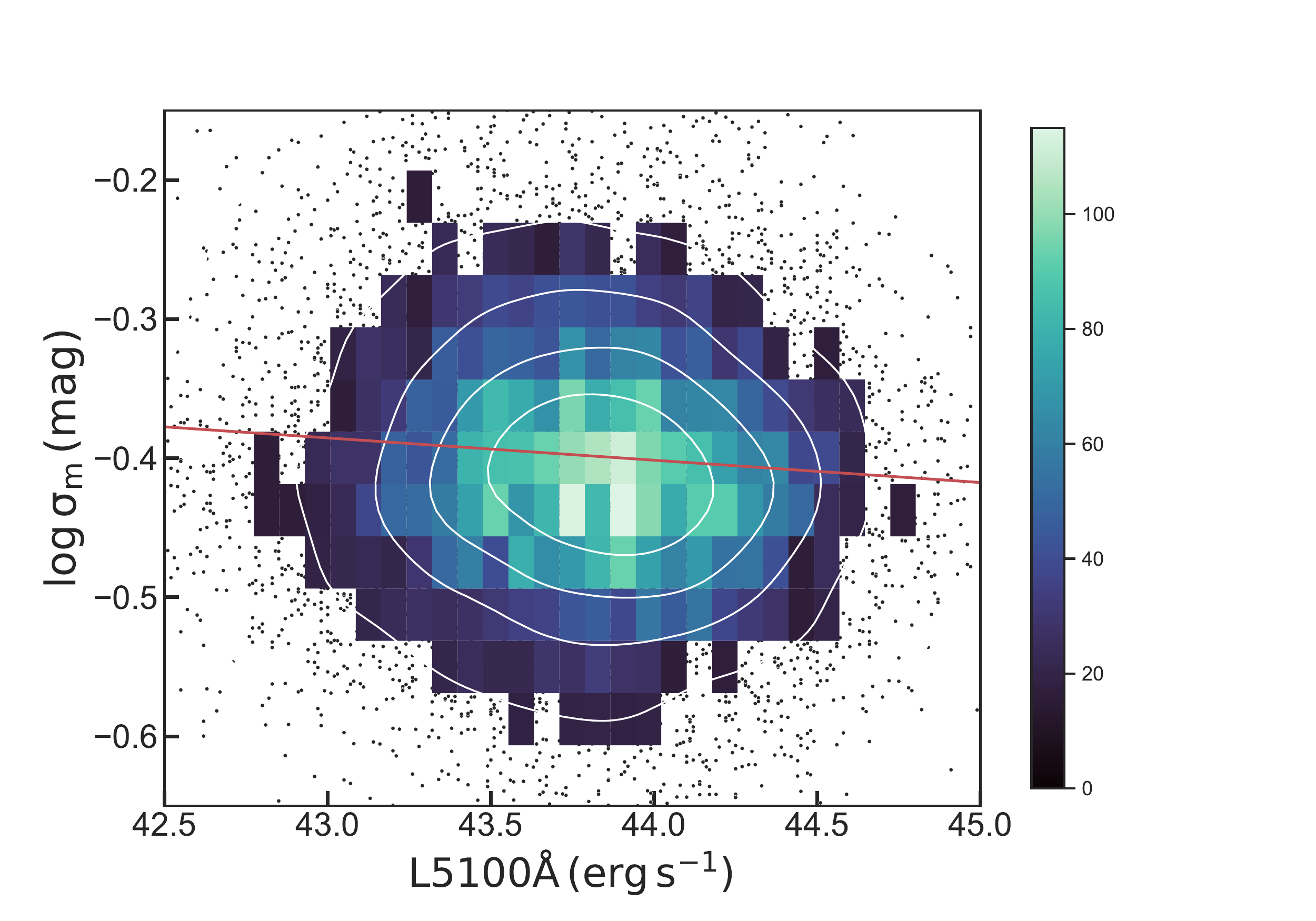}  
\includegraphics[width=0.25\linewidth]{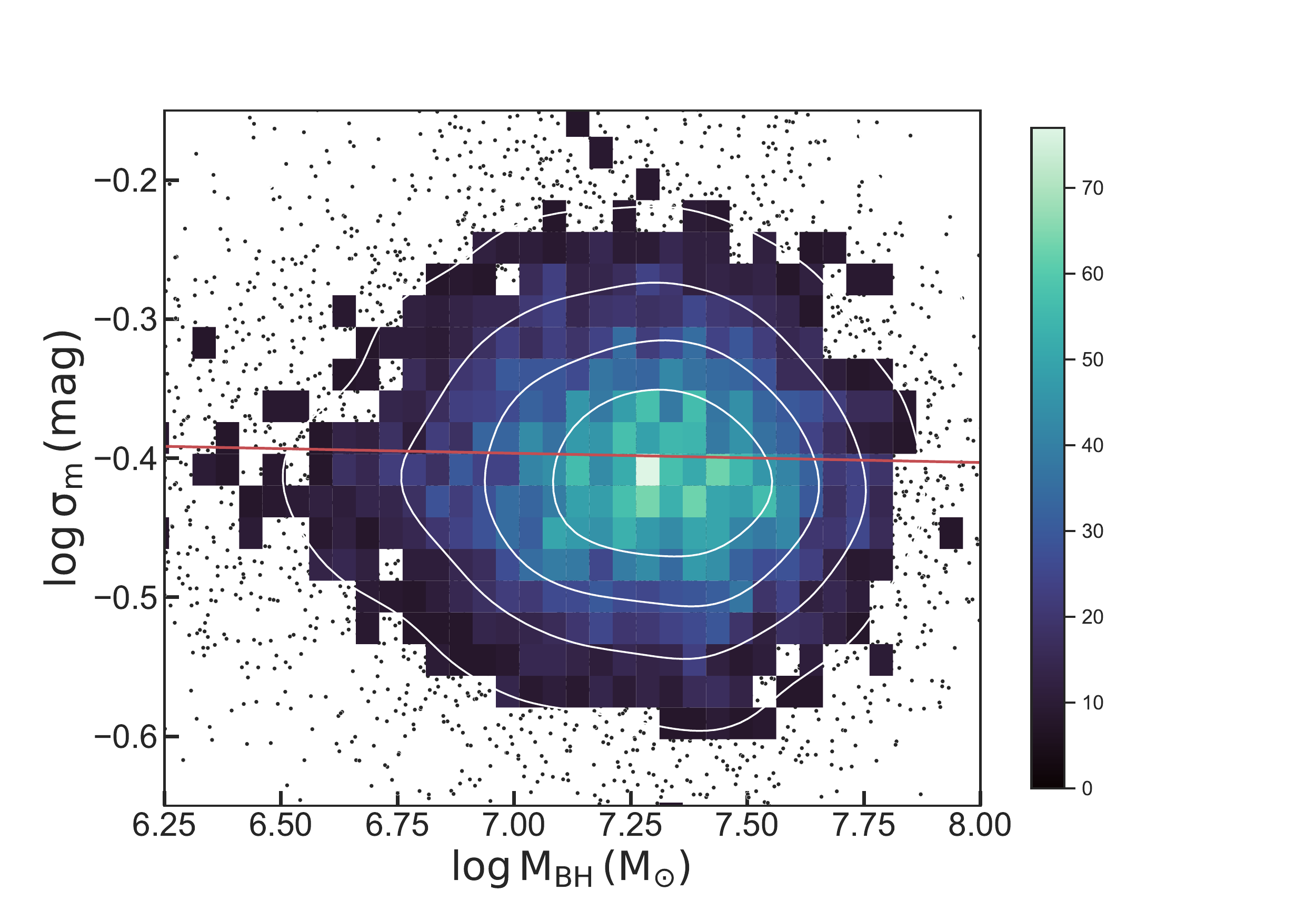}   
\includegraphics[width=0.25\linewidth]{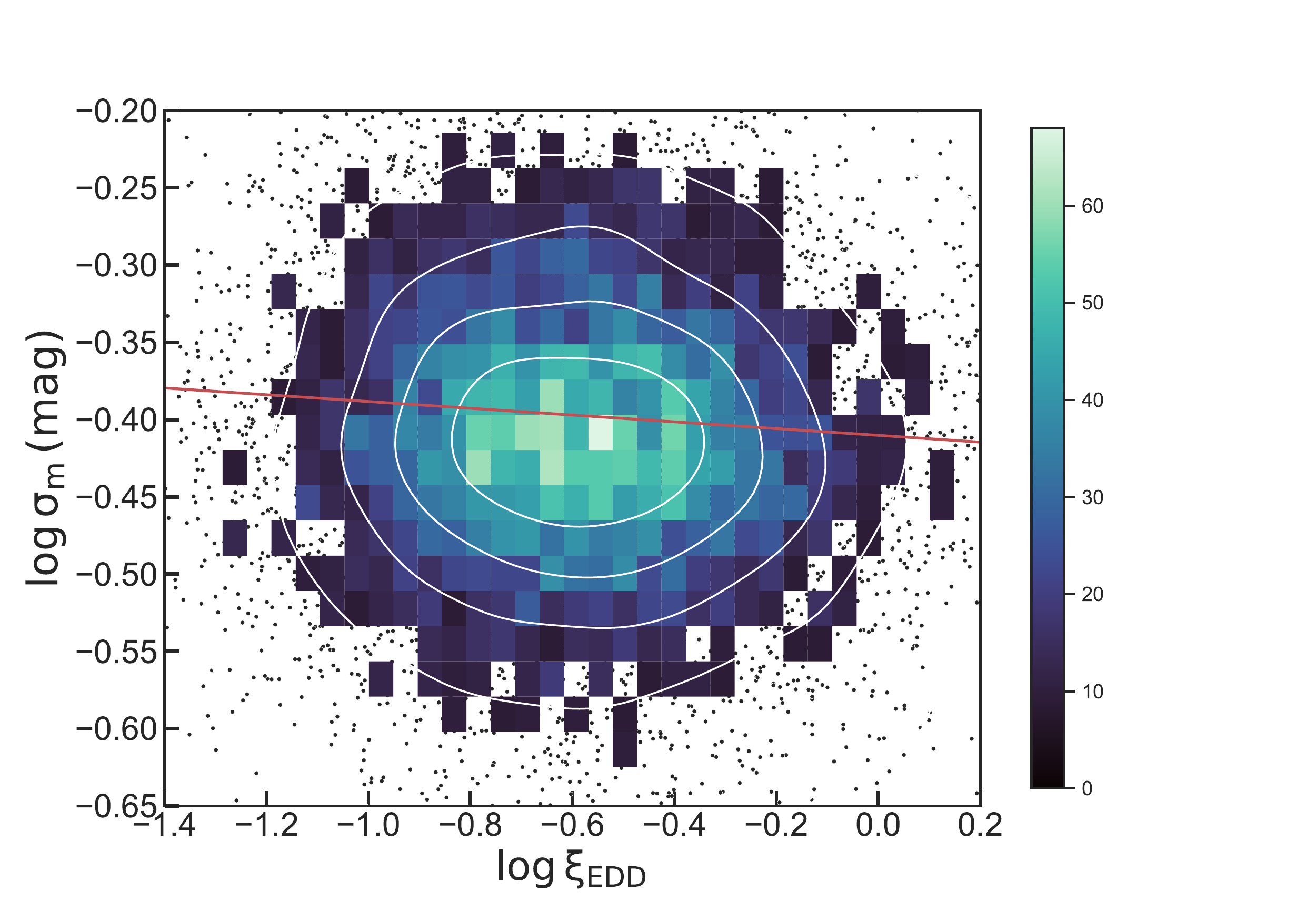}  
\includegraphics[width=0.25\linewidth]{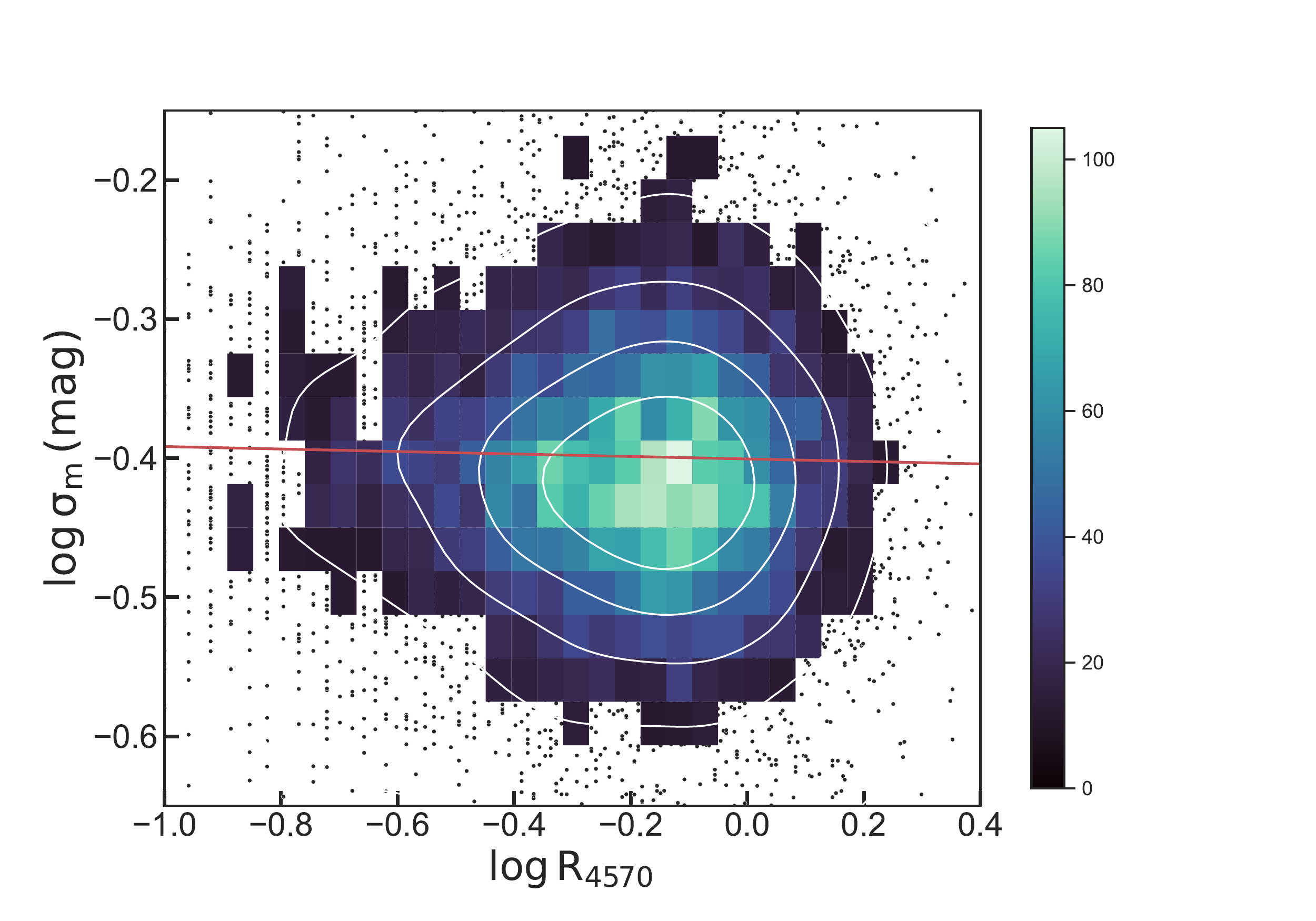}   
\includegraphics[width=0.25\linewidth]{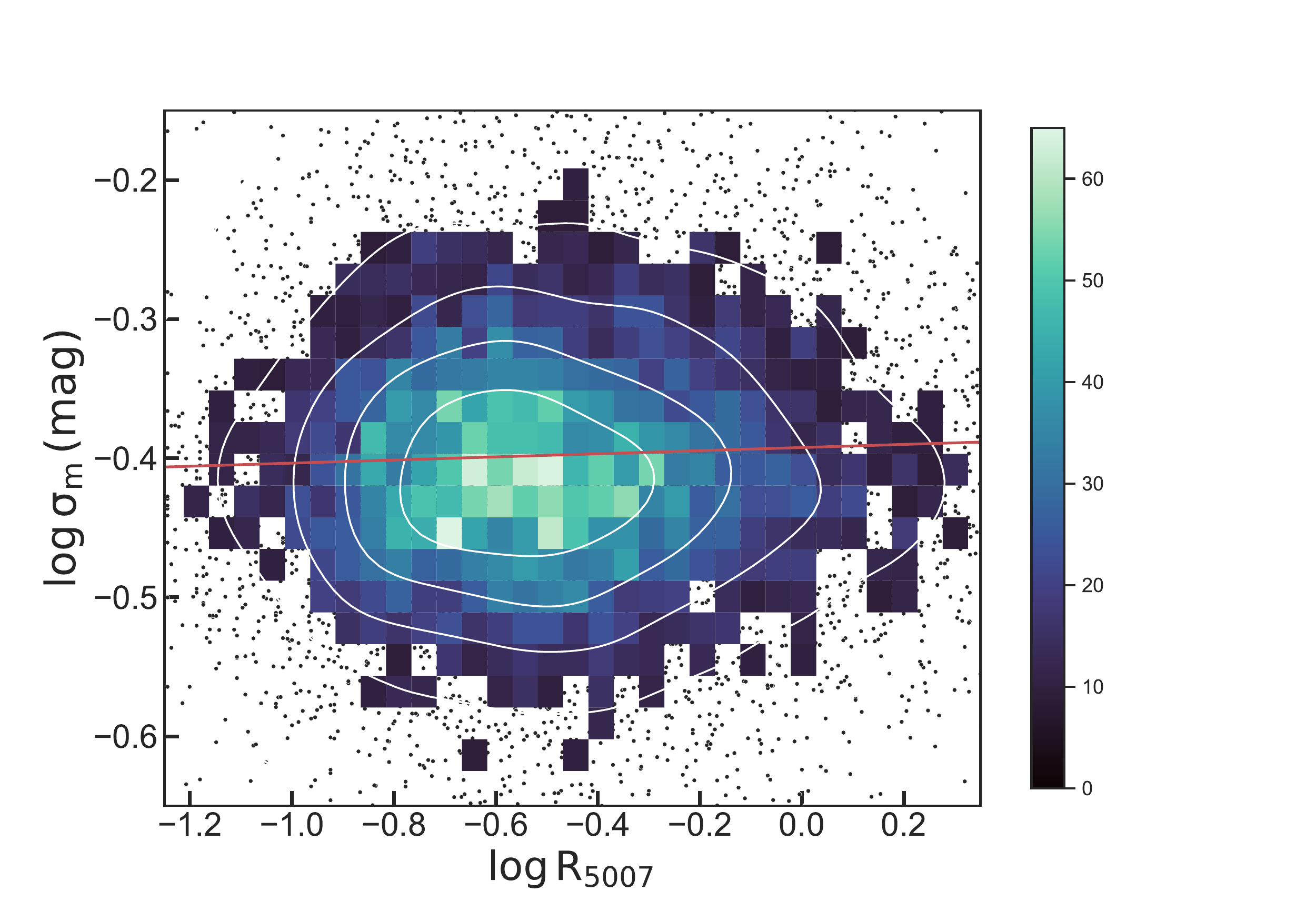}   
\caption{ The relation between $\sigma_m$ and luminosity in 5100 \AA, black hole mass, Eddington ratio, R4570 as well as R5007 in NLS1 galaxy sample. }   
\label{fig:linescan}
\end{figure*}


\begin{table*}
\caption{ The relation between $\sigma_m$ and luminosity, black hole mass, Eddington ratio, R4570 as well as R5007 in NLS1 galaxy sample. } \label{Tab.XXX1}
\centering
\begin{tabular}{|c|c|c|c|}
\hline \hline
  & { Spearman correlation coefficient } & { p value } & { expression }     \\ 
\hline
$\rm\sigma_m-L5100\AA  $  &  -0.060   &  $3.981\times10^{-20}$   &  $ -0.016 x + 0.303 $   \\
$\rm\sigma_m-M_{BH}$      &  -0.065    &  $2.906\times10^{-10}$   &  $ -0.047x- 0.592 $      \\
$\rm\sigma_m-\xi_{Edd}$   &  -0.059    &  $1.049\times10^{-8}$    &  $ -0.051x - 0.960 $     \\
$\rm\sigma_m-log R4570 $  &  -0.023    &  $3.613\times10^{-23}$   &  $-0.009 x - 0.4005 $     \\
$\rm\sigma_m-log R5007 $  &   0.029     &  $3.217\times10^{-3}$    &  $0.011 x - 0.392  $     \\
\hline
\end{tabular}
\end{table*}

\subsection{ The variability characteristics of radio sub-sample  }   
In order to understand the influence of radio emission on the observed optical flux variations, we construct a radio sub-sample of 555 objects by crossing match the NLS1 galaxy sample with the FIRST Survey. We get ride of the sources with zero variability variation. Ultimately, 331, 320, 306, 230 and 173 objects are left in g, r, i, z and y bands,  respectively. 

\subsubsection{ Comparison with the amplitude of radio loud and radio quiet samples }   
 We perform a comparative analysis of the variability amplitude between radio quiet and radio loud NLS1 galaxies. The results of cumulative distribution are shown in Figure 4. The variability amplitude of the radio loud  sub-sample is a slightly larger than that in radio quiet sub-sample in g band. No obvious difference were found between the variability amplitude of radio loud and radio quiet NLS1 galaxy samples in r,i z and y bands, respectively.    
 
\begin{figure*}   
\centering 
\includegraphics[width=0.25\linewidth]{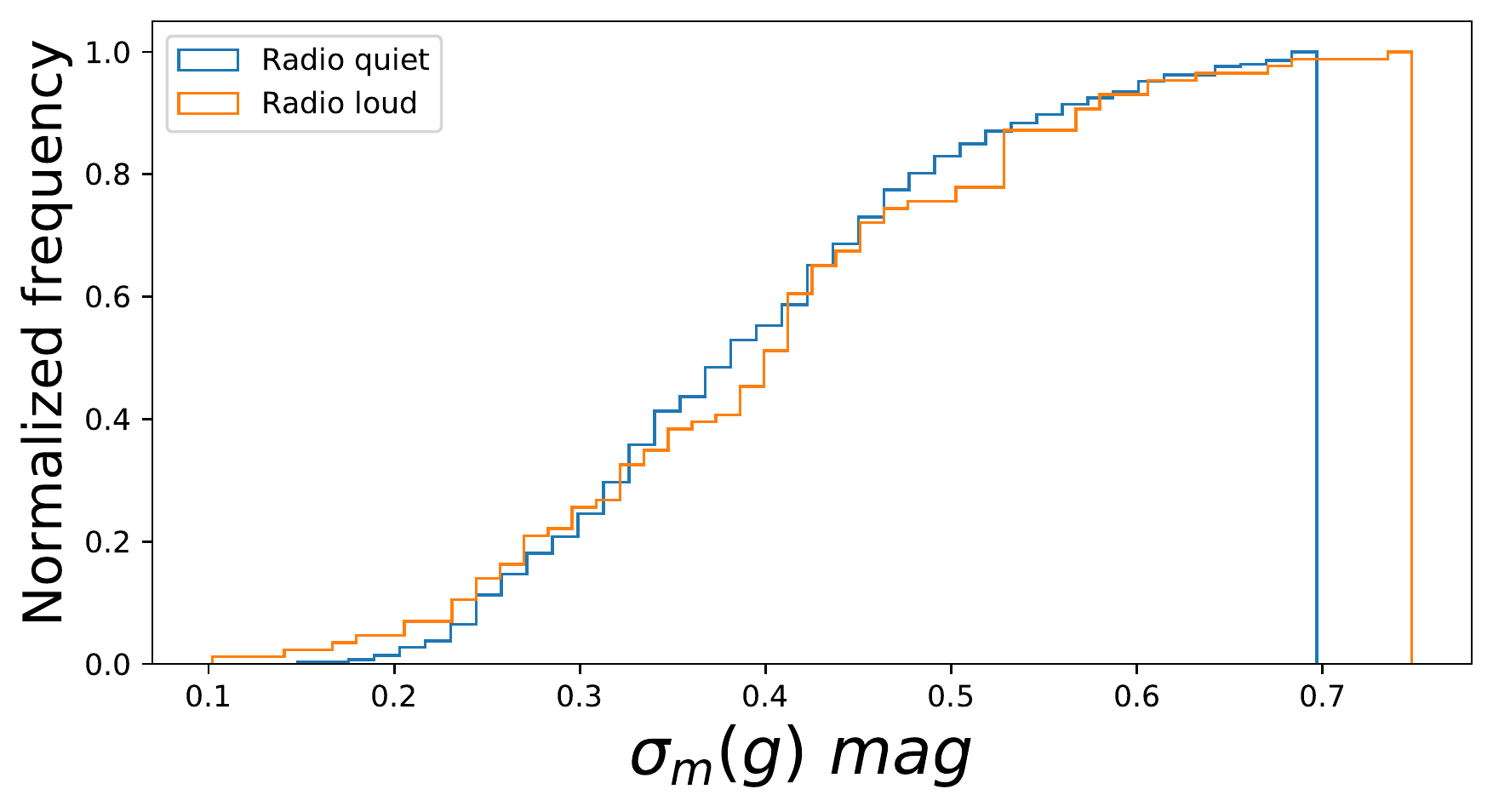}
\includegraphics[width=0.25\linewidth]{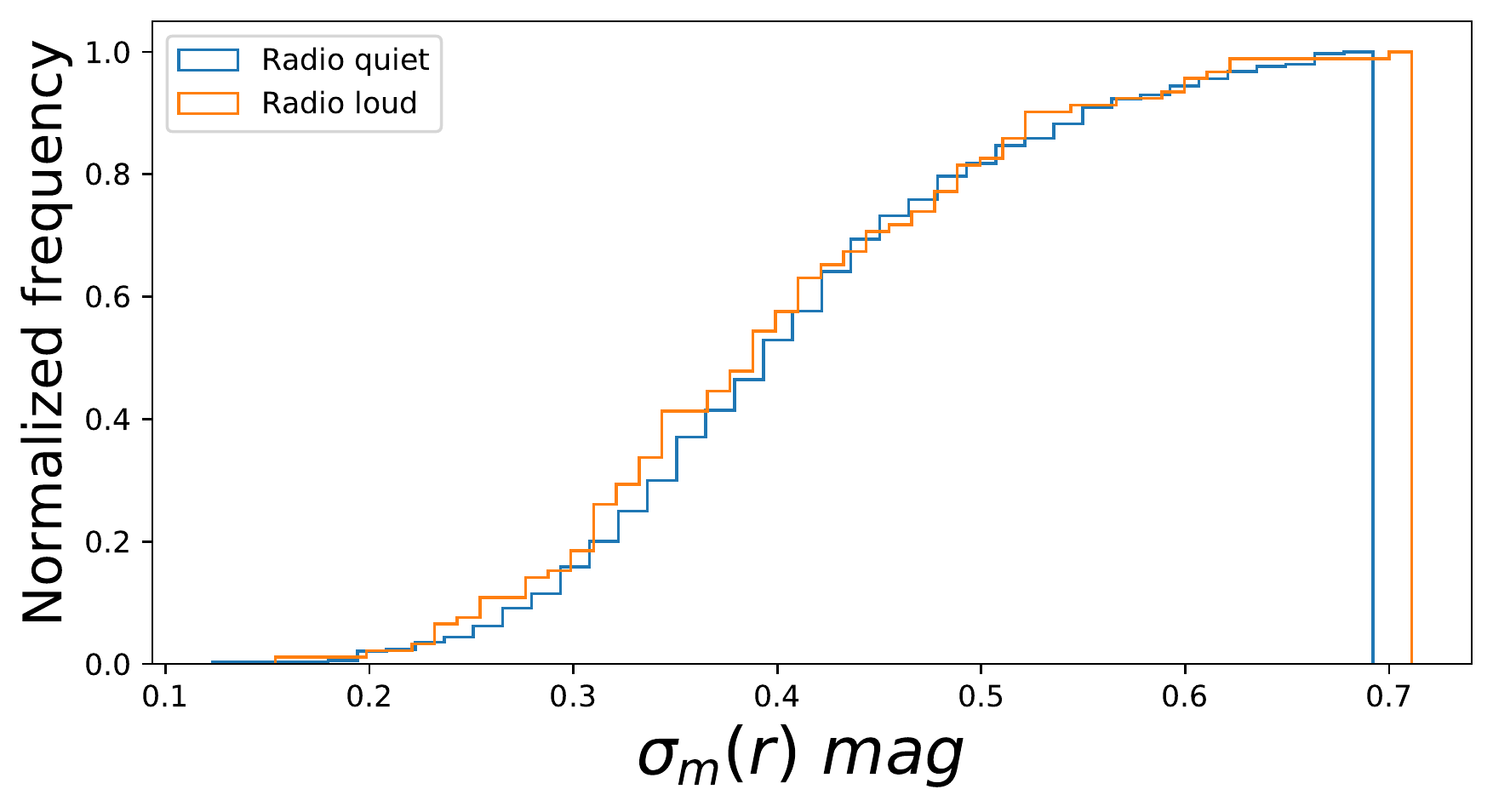} 
\includegraphics[width=0.25\linewidth]{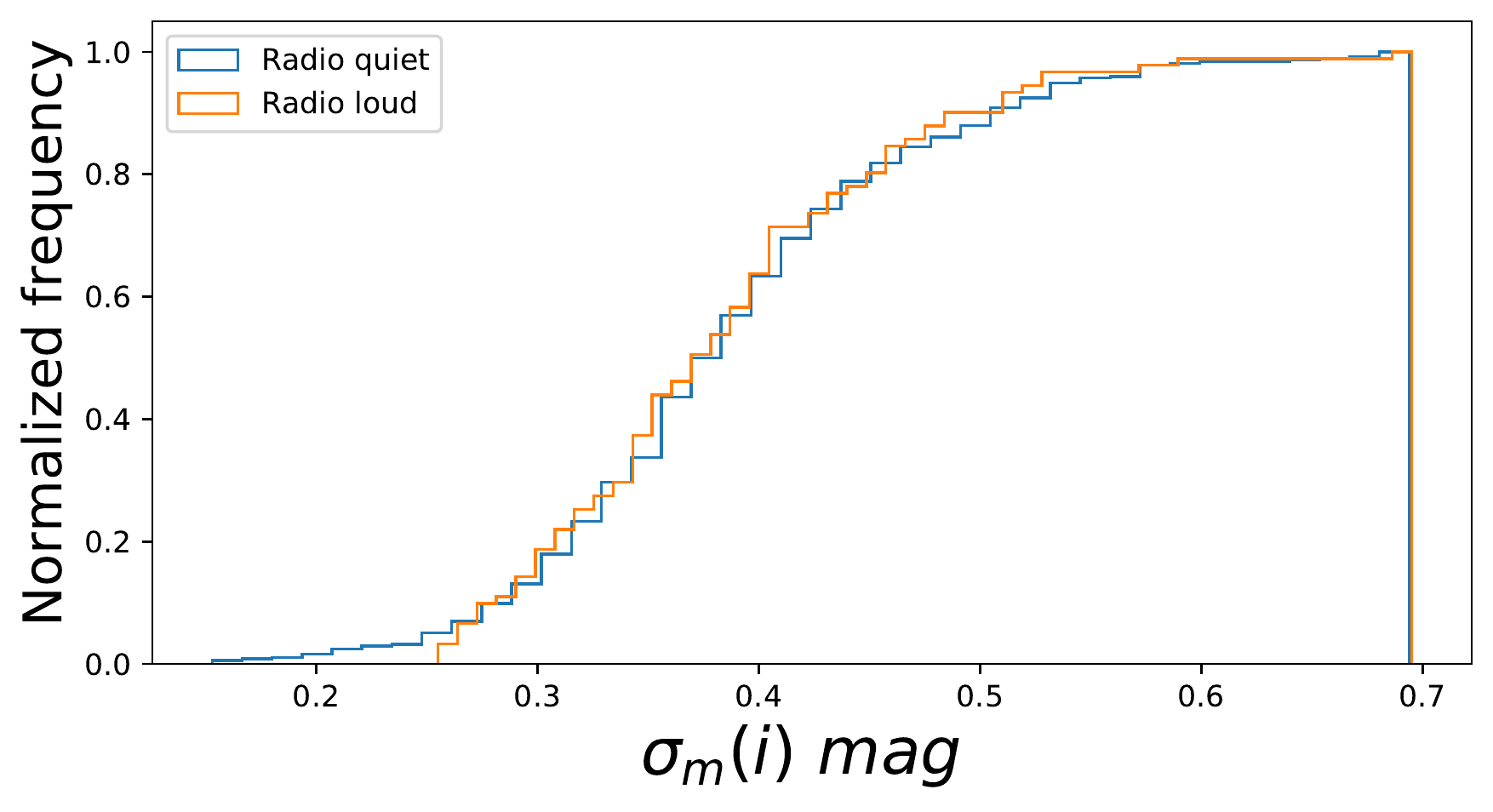}
\includegraphics[width=0.25\linewidth]{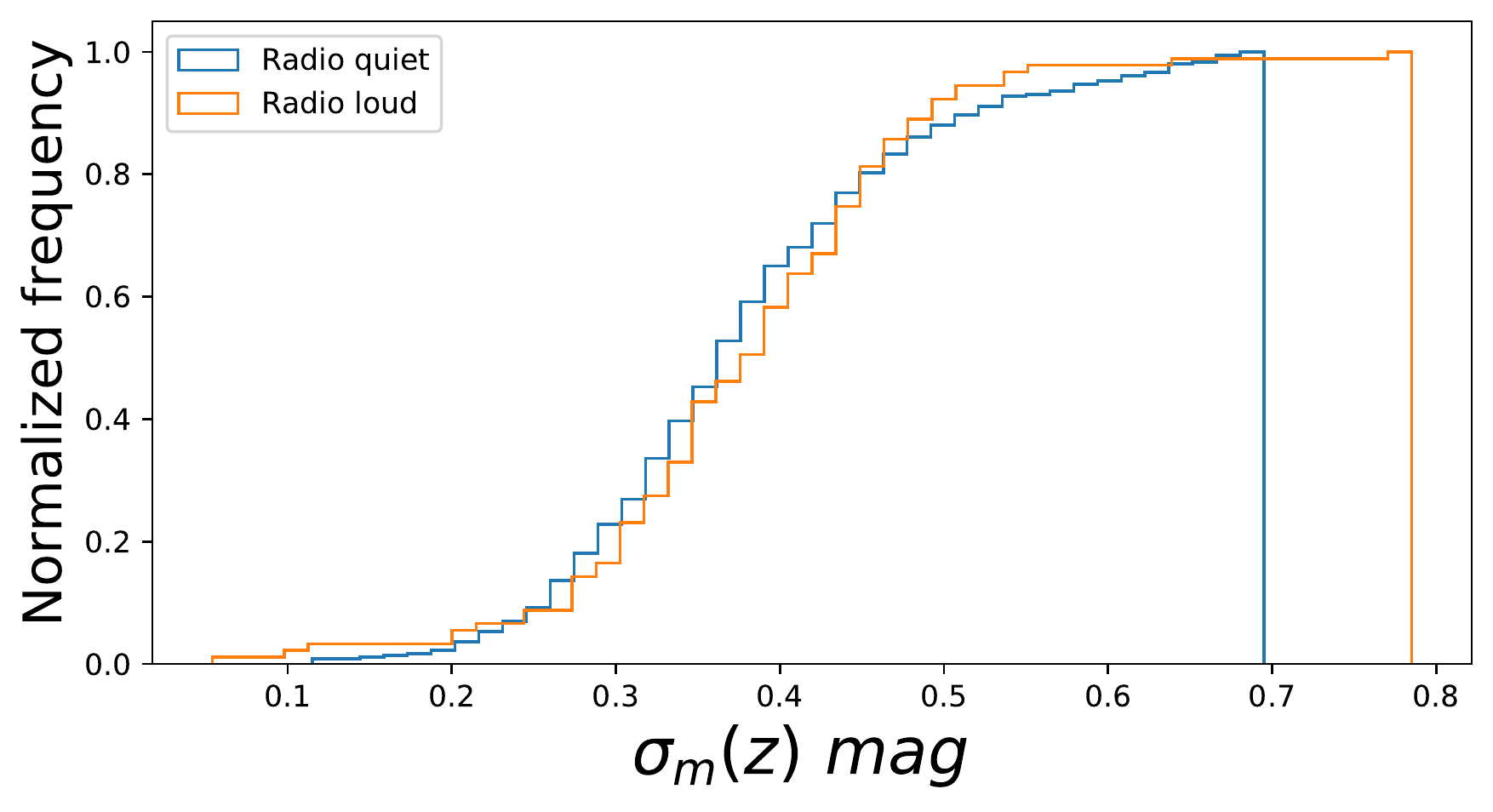}
\includegraphics[width=0.25\linewidth]{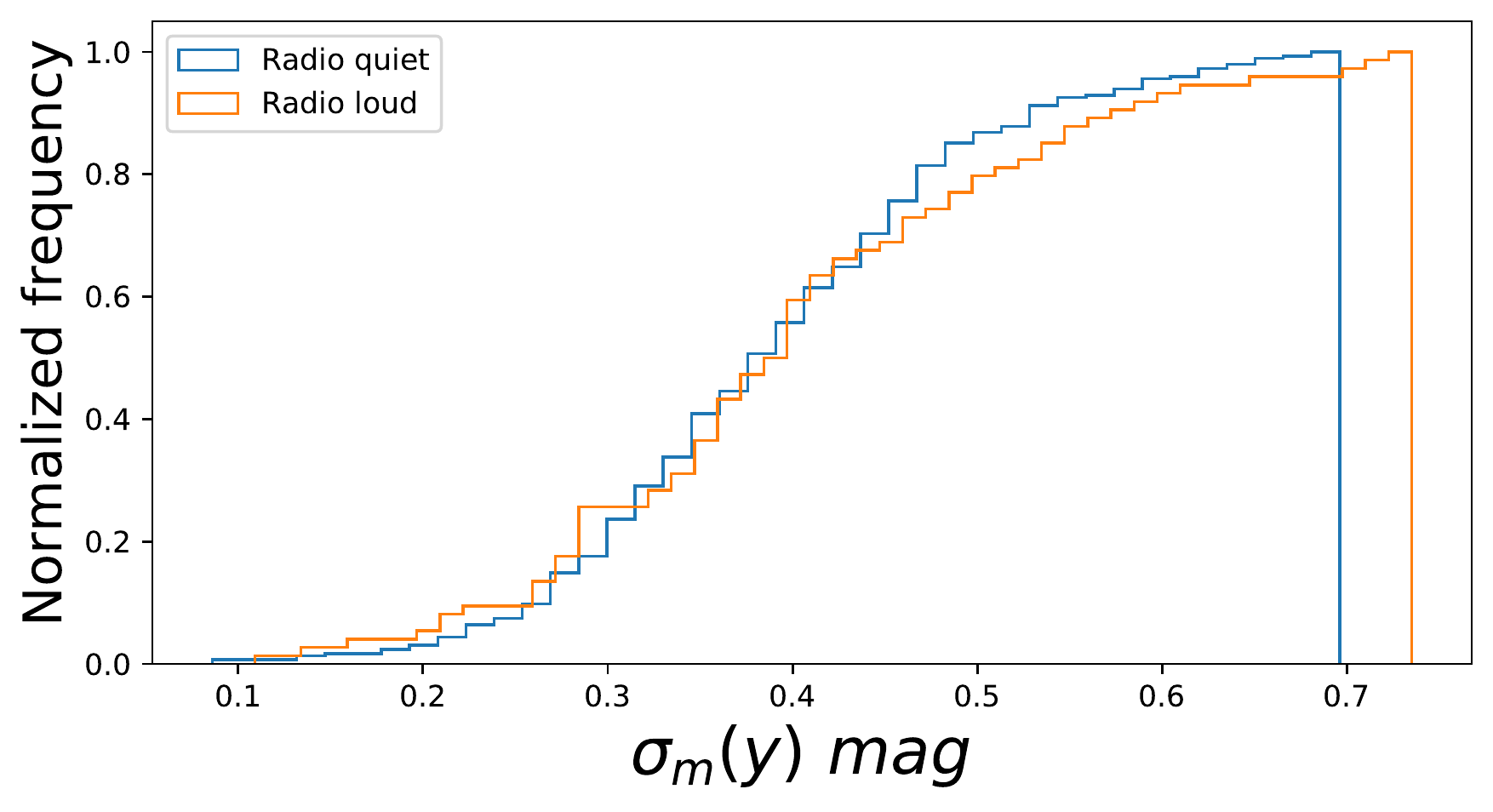}
\caption{ The cumulative $\sigma_m$ and radio loudness in NLS1 galaxies. The upper panel is the  results in g band, r band and i band, respectively. The lower panel is the results in z band and y band, respectively. }
\label{fig:linescan}  
\end{figure*}   

\subsubsection{The relation between variability amplitude and radio loudness }

\citet{2017ApJ...842...96R} found the variability amplitude of radio loud NLS1 galaxies in optical band is higher than that in radio quiet NLS1 galaxies, which might be due to different physical processes in the two classes. They supposed that the optical emission in radio loud objects originated from the presence of both the non-thermal emission from the relativistic jet and the thermal emission from the accretion disk. However, the optical emission was only due to the thermal emission from the accretion disk in radio quiet objects. 

  The relation between variability amplitude and radio loudness is shown in Figure 5. It presents a weak positive correlation in the upper panel (g and r bands) of Figure 5 which is consistent with the reuslts in \citet{2017ApJ...842...96R}. However, no obvious relation was found in the other panel (i, z and y bands) of Figure 5.   

In order to further verify the influence of radio emission on optical variability, we investigate the relationship between variability amplitude in g,r,i,z and y bands and luminosity at 1.4 GHz which has been revised by  K-correction. The results are shown in Figure 6. It is shown a weak positive correlation in the upper left (g band) and middle (r band) panel of Figure 6, but no obvious results in the other band. We speculate that it may be due to the proximity in g and r bands to the fraction of continuum radiation in $\rm 5100 \AA$ of the host galaxy contamination. 
The photometric value in i,z and y band are farther away from $\rm5100 \AA$ band, which may lead to a larger correction error and thus the results become insignificant.   

\begin{figure*} 
\centering
\includegraphics[width=0.25\linewidth]{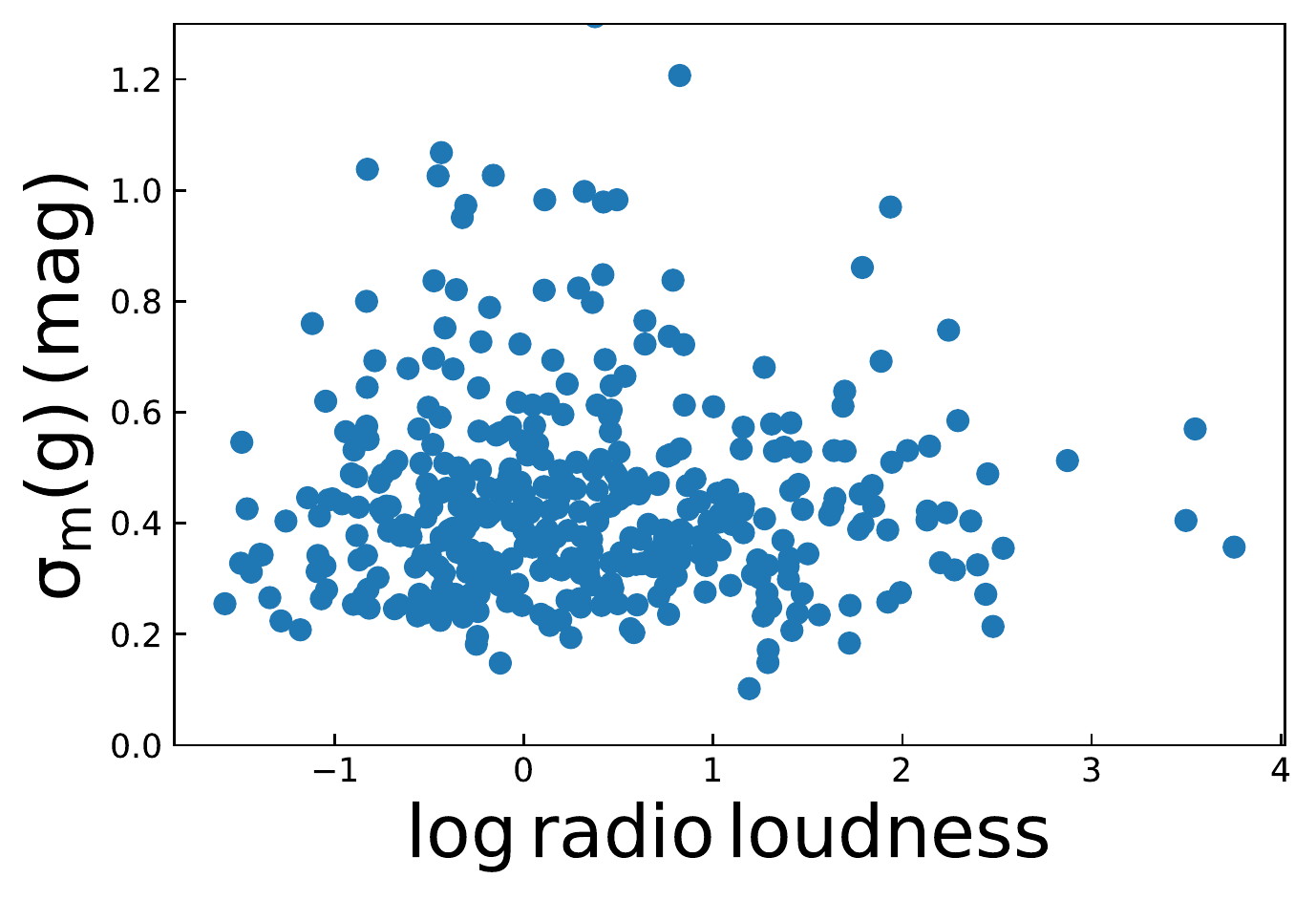}
\includegraphics[width=0.25\linewidth]{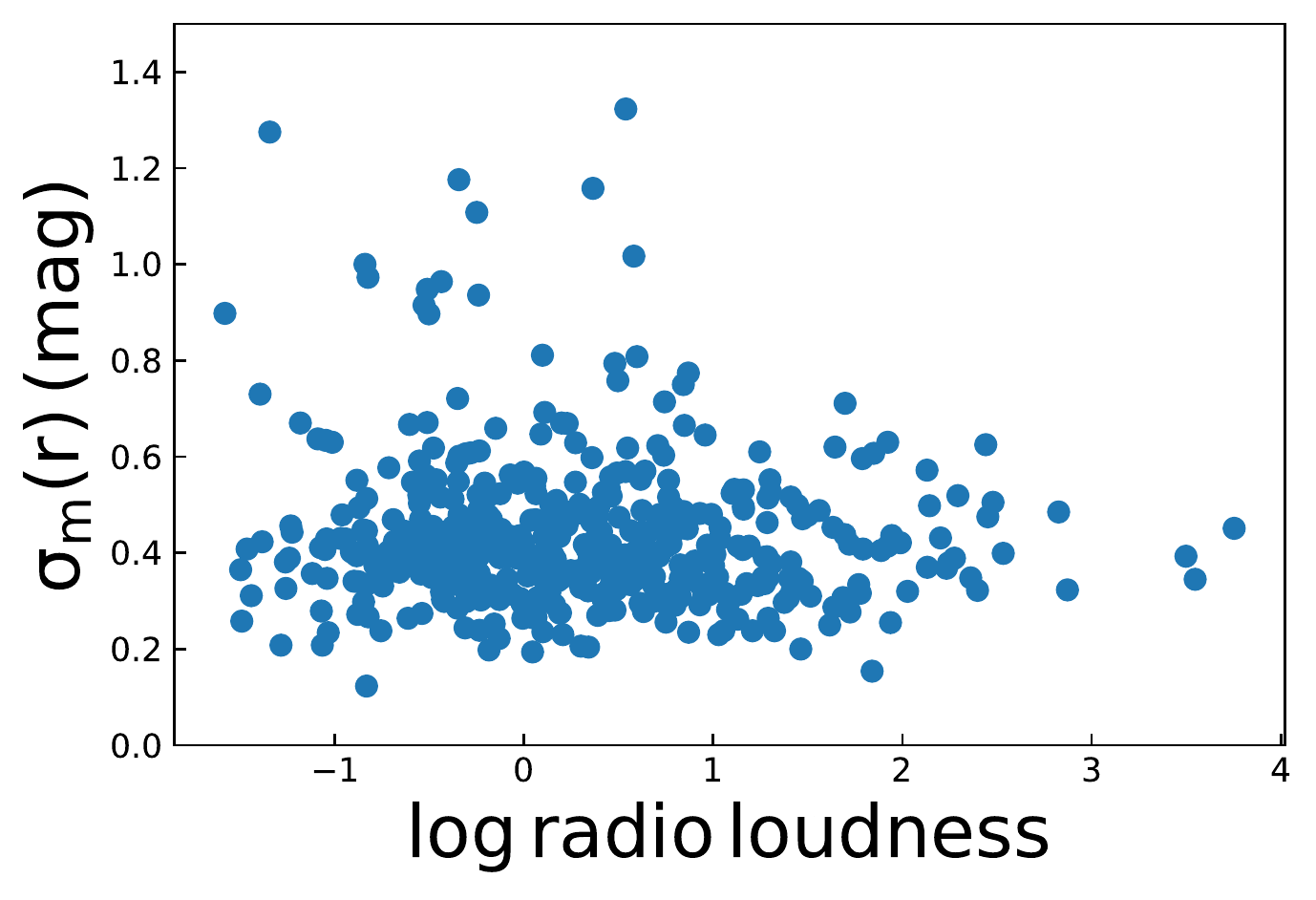}
\includegraphics[width=0.25\linewidth]{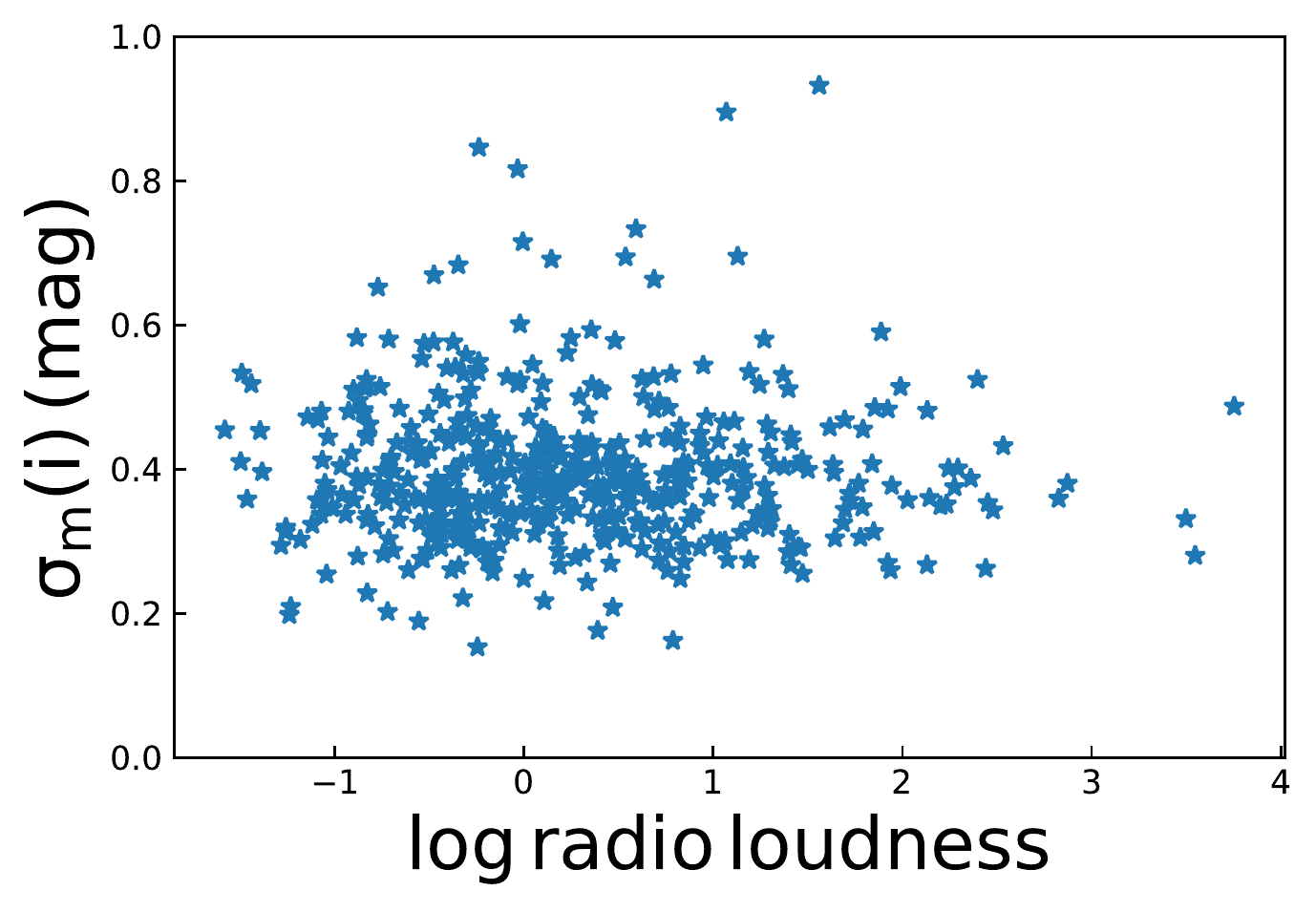}
\includegraphics[width=0.25\linewidth]{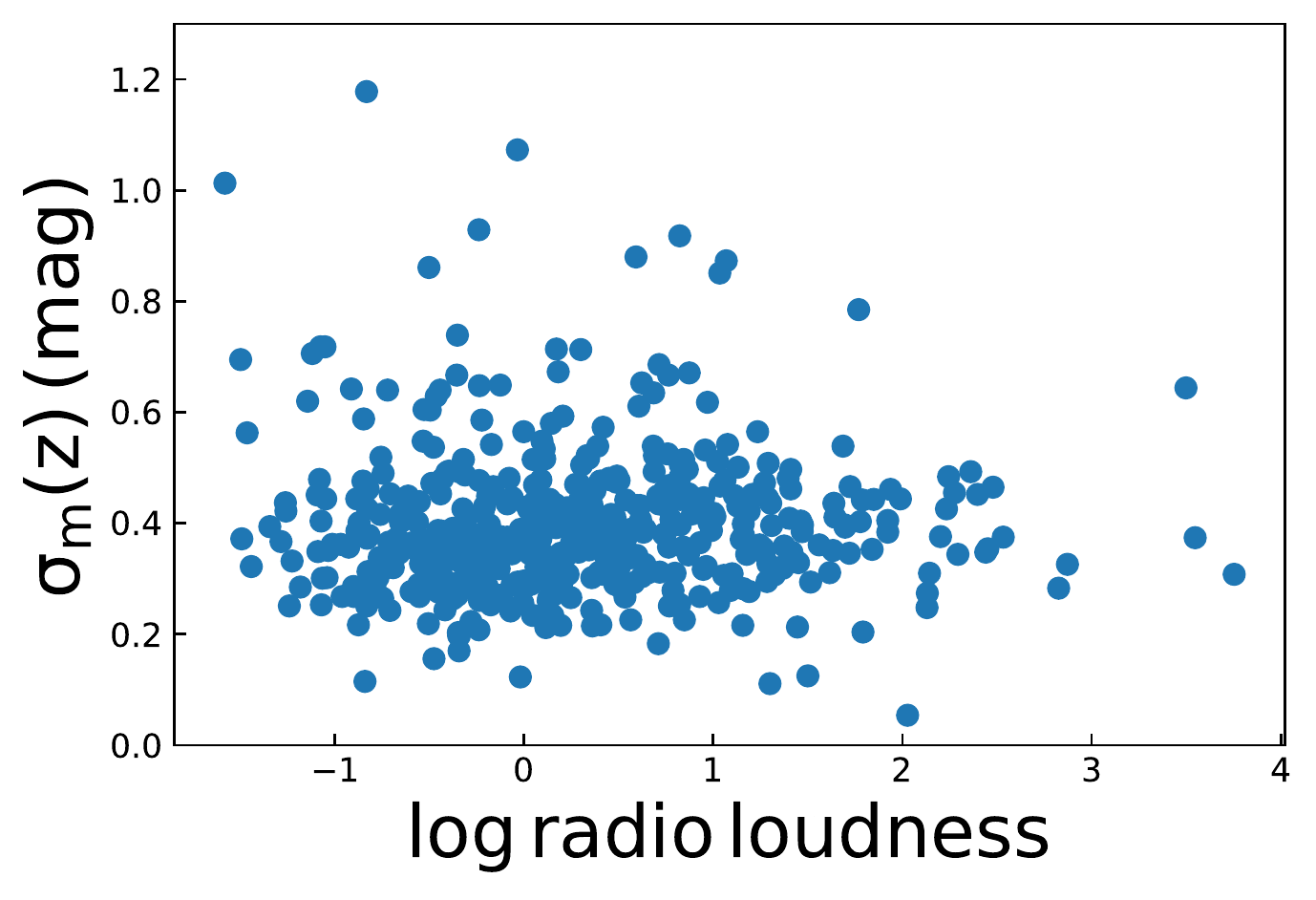}
\includegraphics[width=0.25\linewidth]{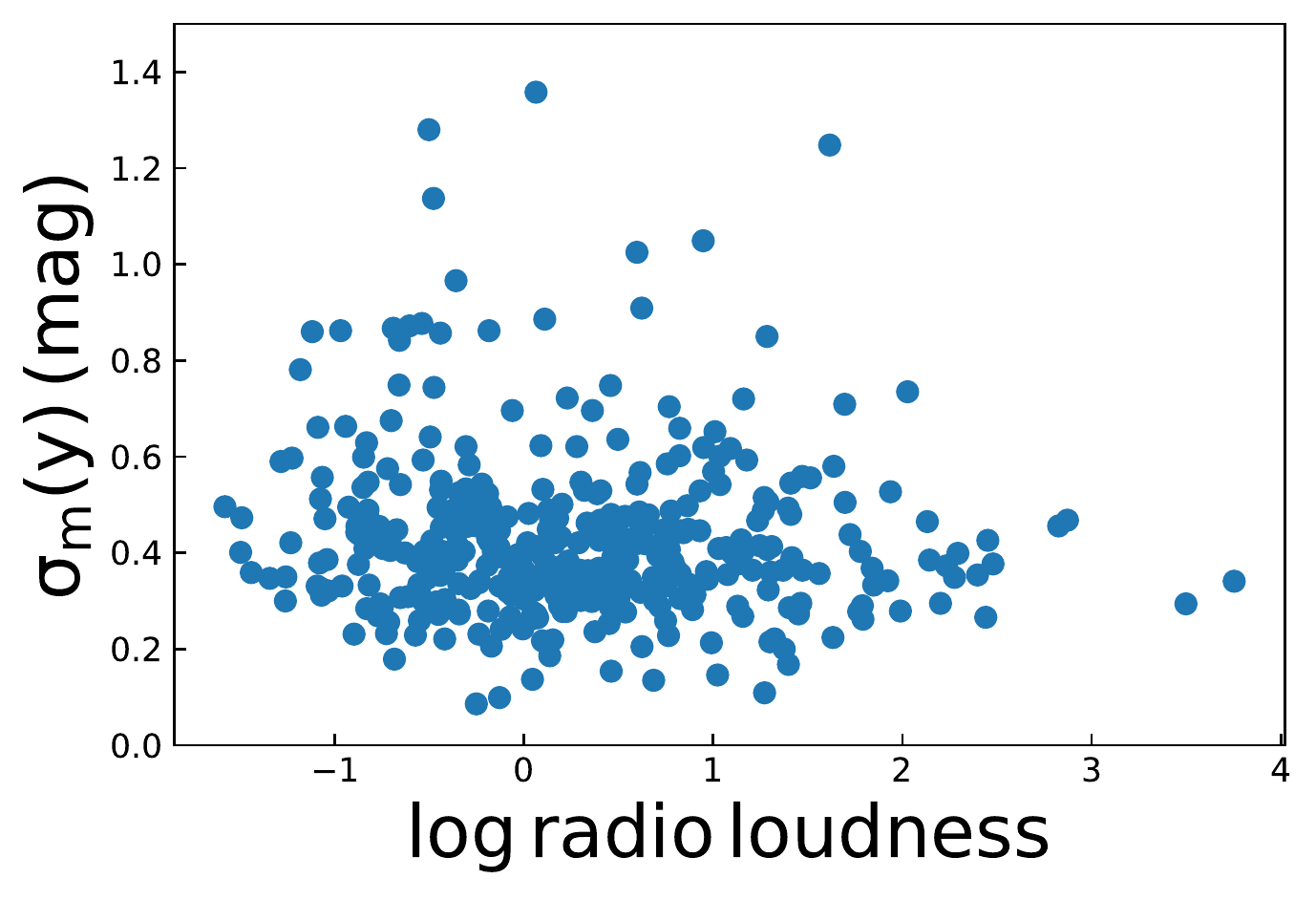}
\caption{ The relation between $\sigma_m$ and radio loudness in NLS1 galaxies. The upper panel is the results in g, r and i bands, respectively. The lower panel is the results in z and y bands, respectively.  }
\label{fig:linescan}
\end{figure*}


\begin{figure*}
\centering 
\includegraphics[width=0.25\linewidth]{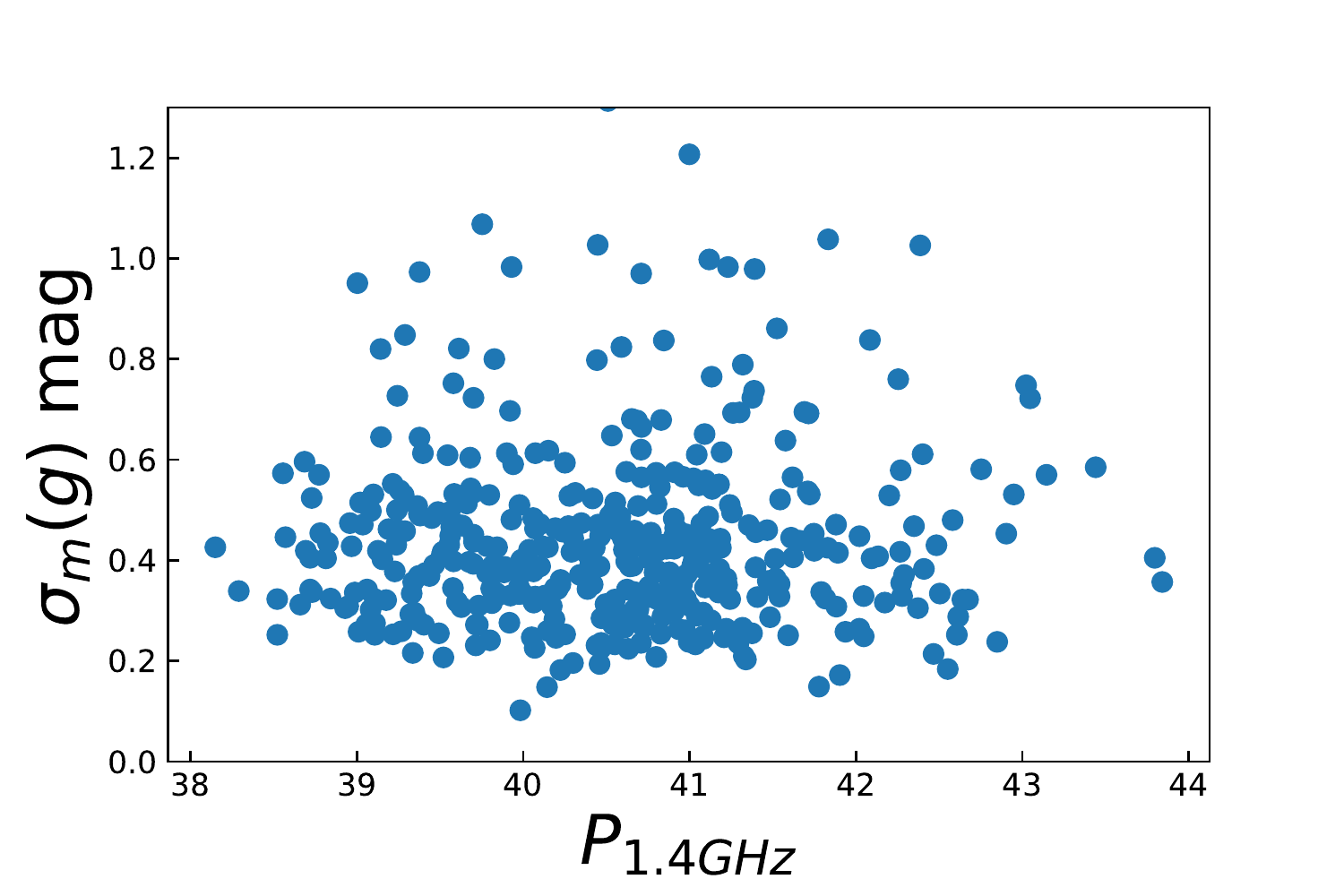}
\includegraphics[width=0.25\linewidth]{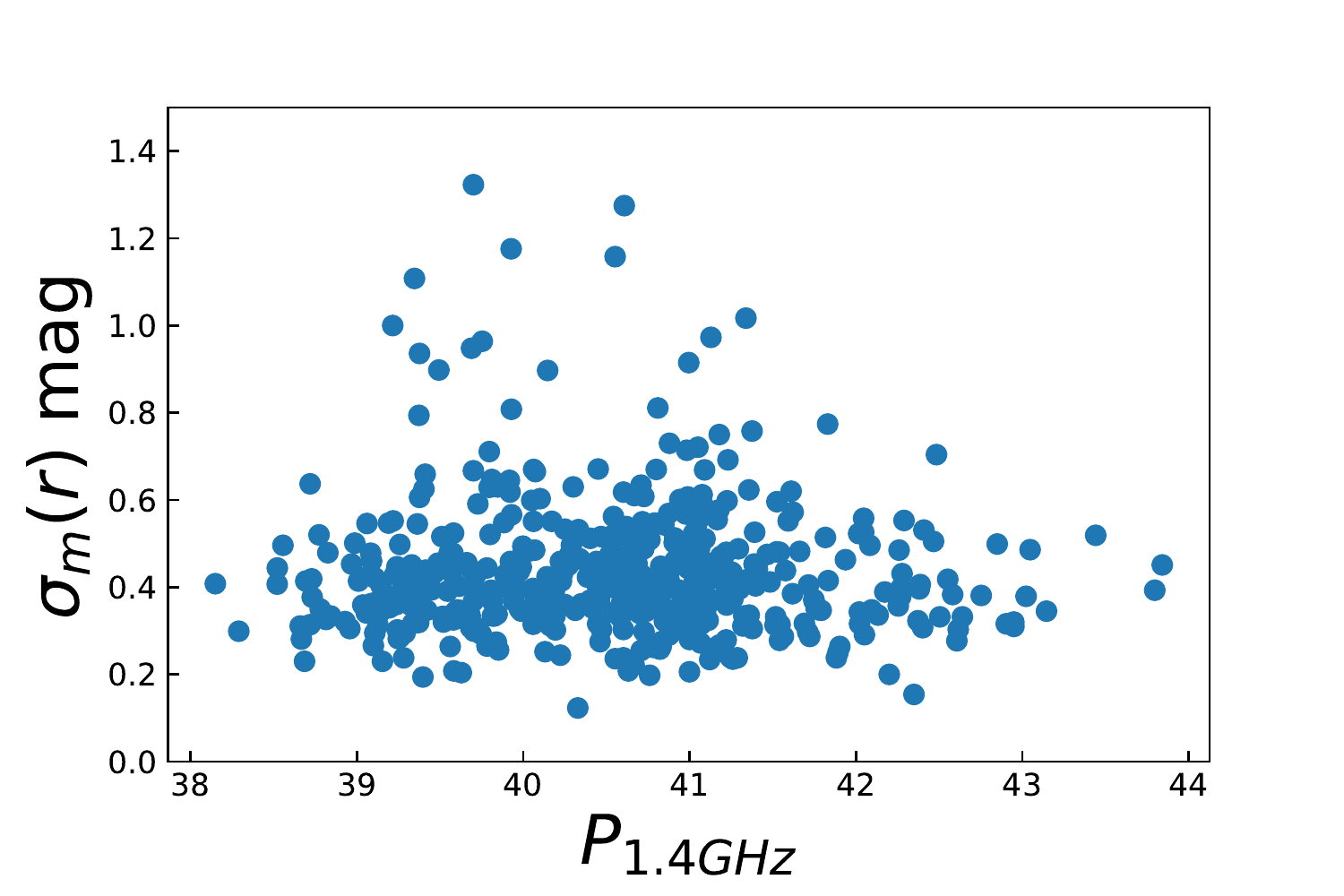}
\includegraphics[width=0.25\linewidth]{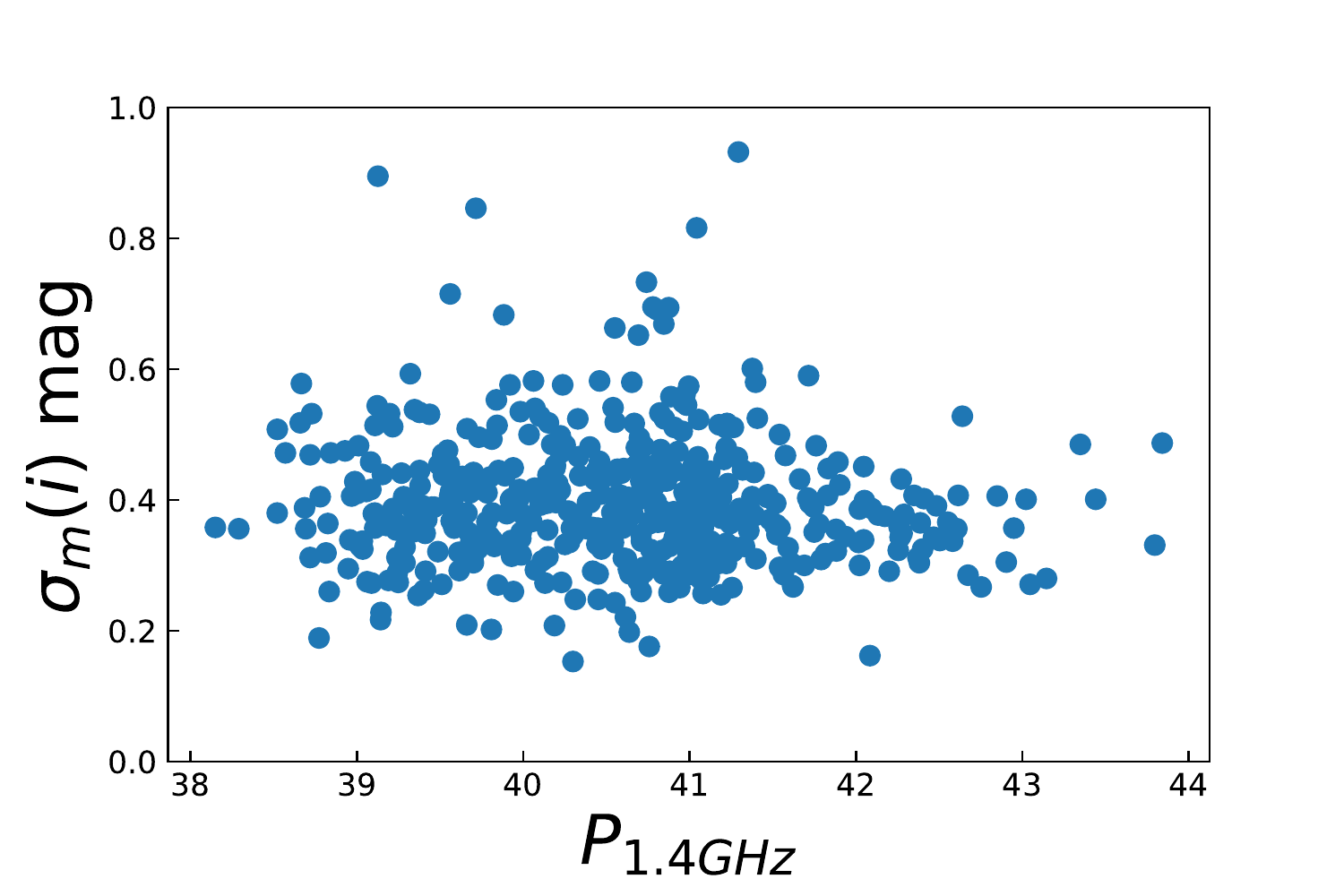}
\includegraphics[width=0.25\linewidth]{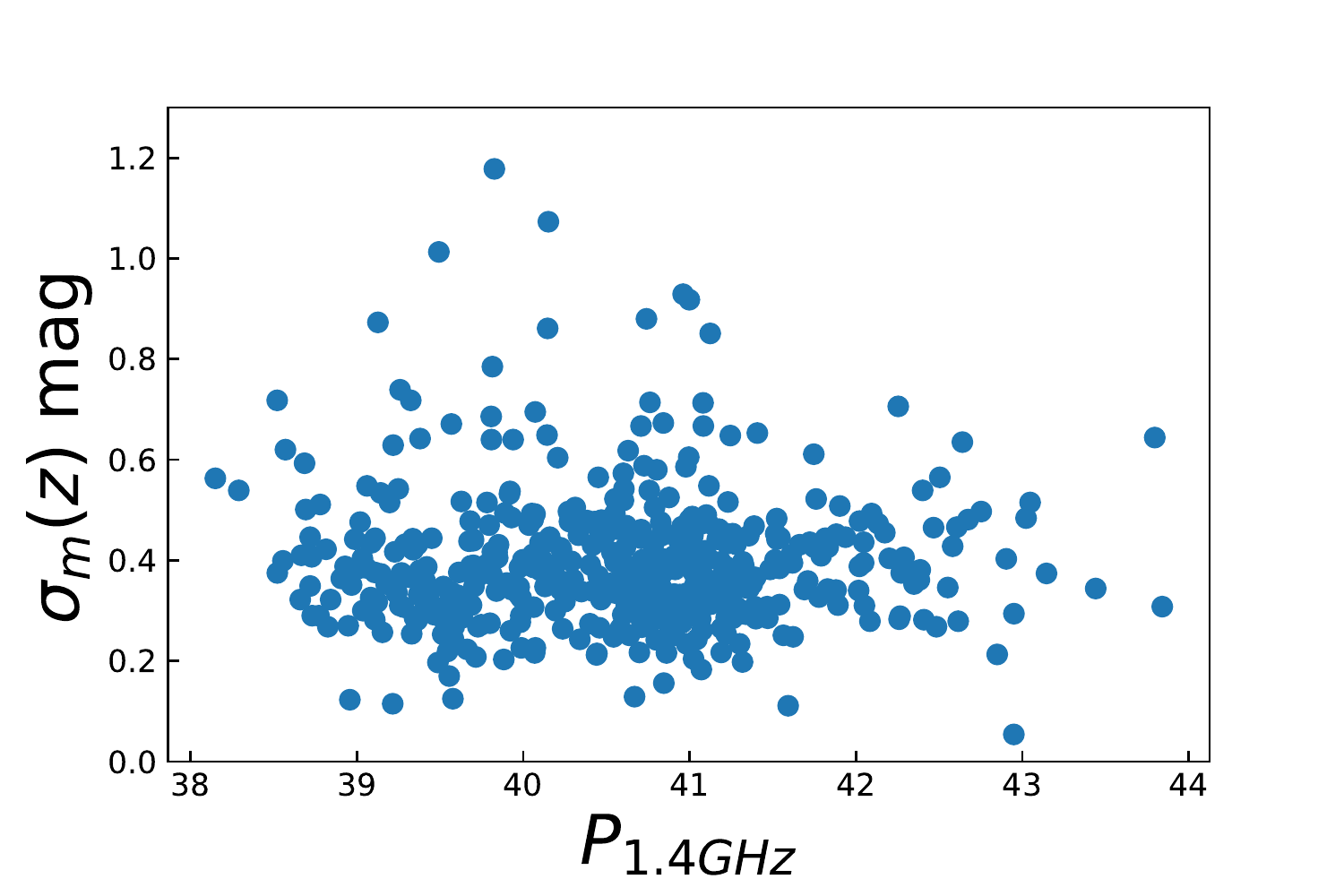}
\includegraphics[width=0.25\linewidth]{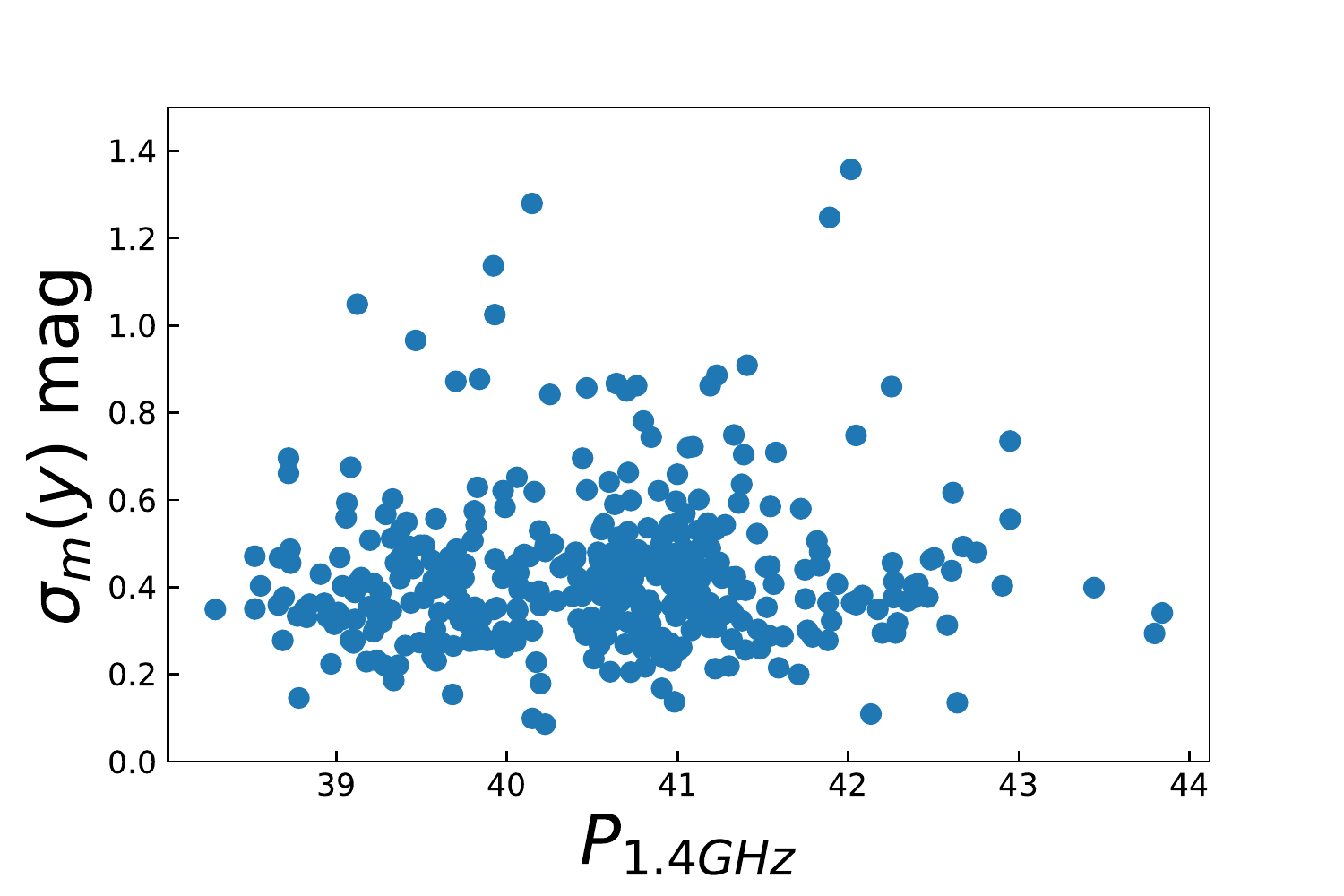}
\caption{ The relation between $\sigma_m$ (g,r,i,z,y) and radio luminosity at 1.4 GHz in NLS1 galaxies. The upper panel is the results in g band,r band and i band, respectively. The lower panel is the results in z band and y band, respectively. }
\label{fig:linescan} 
\end{figure*}


\section{Conclusions}
In this work, we systematically investigate the relationship between optical variability and many physical parameters for 11101 NLS1 galaxies by the data sets from the Pan-STARRS1 survey. The results are summarized as follows.

(1) We investigate the relationship between variability amplitude and absolute magnitude in g,r,i,z and y bands. The results show significant anti-correlations which are consistent with the results in previous works. 

(2) The relationship between optical variability and many physical parameters (e.g., $\lambda L(5100\rm\AA)$, black hole mass, Eddington ratio, $R_{4570}$ and $R_{5007}$) is further analyzed. The results show significant anti-correlation with $L(5100\rm\AA)$, $M_{\rm BH}$, Eddington ratio and $R_{4570}$, but positive correlation with $R_{5007}$. The relation could be explained by the simple standard accretion disk model. 

(3) The relationship between optical variability and radio luminosity/radio loudness was analyzed. The results between optical variability and radio luminosity/radio loudness show weak positive correlation in g and r bands, insignificant correlation in i, z and y bands. The large error of the approximate fraction of host galaxy in i, z and y bands may induce the insignificant correlations.

\begin{acknowledgements}
We thank Zhang Xueguang from Nanjing Normal University, Guo Xiaotong from Nanjing University and Lu Kaixing from Yunnan observatory for the helpful comments and suggestions that have improved this manuscript. This work is funded by the the Fundamental Research Funds for the Universities in Hebei province(No: JYQ202003)

\end{acknowledgements} 

\bibliographystyle{raa}
\bibliography{bibtex}




\end{document}